\documentclass[sn-mathphys,Numbered,iicol]{sn-jnl}

\usepackage{graphicx}%
\usepackage{multirow}%
\usepackage{amsmath,amssymb,amsfonts}%
\usepackage{amsthm}%
\usepackage{mathrsfs}%
\usepackage[title]{appendix}%
\usepackage{xcolor}%
\usepackage{textcomp}%
\usepackage{manyfoot}%
\usepackage{booktabs}%
\usepackage{algorithm}%
\usepackage{algorithmicx}%
\usepackage{algpseudocode}%
\usepackage{listings}%

\theoremstyle{thmstyleone}%
\theoremstyle{thmstyletwo}%

\theoremstyle{thmstylethree}%

\begin{document}

\title[Localization of matters coupled nonminimally to gravity on scalar thick branes]{Localization of matters coupled nonminimally to gravity on scalar thick branes}


\author*[1]{\fnm{Muhammad Taufiqur} \sur{Rohman}}\email{m.taufiqur25@gmail.com}

\author[1]{\fnm{} \sur{Triyanta}}\email{triyanta@itb.ac.id}

\author[1]{\fnm{Agus} \sur{Suroso}}\email{suroso@itb.ac.id}

\affil*[1]{\orgdiv{Theoretical High Energy Physics Research Division}, \orgname{Department of Physics, Faculty of Mathematics and Natural Sciences, Institut Teknologi Bandung}, \orgaddress{\street{Jl. Ganesha 10}, \city{Bandung}, \postcode{40132}, \state{Jawa Barat}, \country{Indonesia}}}




\abstract{We are investigating the localization of matter that interacts nonminimally with gravity within thick braneworld models generated by a scalar bulk. Our review focuses on two models of scalar thick branes. The natural mechanism is used to analyze the localization of the fields. Without losing the point of field localization, we examine matter field localizations by considering the asymptotic behavior of the warp function on $z$ towards infinity. Both massless and massive modes of the nonminimally coupled scalar field are localized on the brane in both models. When the coupling is minimal, the scalar field is localized for the massless mode. A nonminimally coupled vector field behaves similarly to the nonminimally coupled scalar field, the massless and massive modes in both models are localized. For a nonminimally coupled spinor field, in model 1, we observe localization of the spinor for massive mode, while in model 2, only the massless spinor field is localized.}

\maketitle

\section{Introduction}\label{sec1}

Braneworld is a higher than 4-dimensional model that assumes the 4-dimensional universe with its matters are confined within a hypersurface called a brane, while gravity is able to propagate in the extra dimension called a bulk. In this term, matter fields in the universe should be localized on the brane. 
The localization of gravity and various bulk matter fields is a crucial issue in braneworld theory. To recover the effective 4-dimensional gravity, the gravity zero mode must be localized on the brane. To build up the standard model, various bulk matter fields should be localized on the brane through a natural mechanism.
The field localization is a mechanism of reducing the 5-dimensional action of the field into the 4-dimensional one, which requires finiteness of the normalization and mass conditions throughout the extra coordinate. This is an important issue in braneworld theory since not all types of fields can be localized.
As an example, some types of fields in the Randall-Sundrum (RS) model \cite{Randall1999, Randall1999a} are not localized on the brane \cite{Bajc2000}. The Modified Randall-Sundrum (MRS) model, which is defined from the RS model through an extra coordinate transformation, on the other hand, shows better localization properties as compared to the RS model \cite{Jones2013, Wulandari2017, Wulandari2019}. 
In fact, scalar, vector, and spinor fields interacting minimally with gravity are localized on the brane in the MRS model \cite{Jones2013, Wulandari2017}. In the RS model, on the other hand, only massless scalar fields are localized on the brane.  

There are two types of brane models, namely thin brane and thick brane models. The RS, MRS, and other RS-like models are example of thin-brane models. The brane is highly ideal because its thickness is minimal. 
The corresponding warp factor leads to the appearance of a Dirac delta function term in the Einstein's equation at points on the brane.  
There are two types of RS models, namely the RS I model which defines the TeV brane (visible brane where all Standard Model particles are confined, at $y = \pi r_c$) and the Planck brane (hidden brane, at $y=0$), which is located in a compactified extra spatial dimension on an $S^1/\mathbb{Z}_2$ orbifold with a radius $r_c$. They are connected exponentially through the warp factor without introducing any new fields beyond the standard model as a solution to the hierarchy problem  \cite{Randall1999a}.  
The other, the RS II model defines the Planck brane at $y = 0$ where the Standard Model particles are assumed to confine and the TeV brane is shifted to position $r_c \rightarrow \infty$ (non-compact extra dimension) \cite{Randall1999a}.
Then, the MRS model is a brane model defined from the RS model through extra coordinate transformation, $dy = e^{-k|z|} dz$.
Besides its better field localization, the finiteness of the proper distance distinguishes the MRS from the RS model \cite{Jones2013}.
The localization properties of the interacting fields defined through a gauge mechanism \cite{Wulandari2017} and through a Yukawa coupling \cite{Wulandari2019} in the MRS model have been discussed.

On the other hand, thick branes are more realistic models that are developed based on the existence of a minimal length scale in fundamental physics called the Planck scale. 
In the thick brane, it is not necessary to provide the junction conditions for the gravitational field equation. The warp factor solution is a smooth function \cite{Liu2018a}.
The thick brane model can be constructed from any 5-dimensional gravity theory, which conducts the smooth warp factor.
From the gravitational field action, a set of field equations can be derived, and the warp factor function can be examined. The solution of field equations in the extra coordinate and the warp function are then utilized for examining the localization of the field. Some thick brane models have been developed, including thick branes generated by pure curvature \cite{Liu2008, Guo2013, Herrera-Aguilar2010a}, scalar bulk \cite{Liang2009, Bazeia2009, Gremm2000a}, vector \cite{Geng2016}, and spinor field \cite{Dzhunushaliev2011,Cui2023}. The thick brane models are not only formulated in the standard theory of gravity but also in modified gravities, such as $f(R)$ \cite{Bazeia2014} and $f(R, T)$ gravities \cite{Bazeia2015, Gu2017, Rohman2021b}, etc. 

The nonminimally coupled field to gravity is an intriguing topic of discussion in quantum field theory and cosmology. As a modification of general relativity, it provides insights into cosmological and theoretical concerns, such as the late-time cosmic acceleration, cosmological inflation with boson as inflaton \cite{Bezrukov2008}, dark energy-dark matter interaction \cite{Harko2014}, and so on. 
Some aspects of the RS model with nonminimally coupled bulk scalar field have come to the attention of some researchers, including stability of the model \cite{Farakos2005, Farakos2006}, impact of the nonminimal coupling constant on the brane structure, and localizations of gravity, scalar, spinor, and vector fields \cite{Guo2012}. Localization of matter fields and implications for the potential in the Schr\"odinger-like equation in the sine-Gordon and double sine-Gordon thick branes are found in \cite{Moazzen2017}.

In recent studies, the localization of a $U(1)$ gauge field that is nonminimally coupled to gravity in RS-like braneworld models has been explored \cite{Zhao2023}. Another study delves into the localization of gauge fields (Abelian and Kalb-Ramond), which are coupled to the scalar field responsible for generating the thick brane and also coupled to torsion \cite{Moreira2023}. The investigation of scalar and vector field localization on a de Sitter brane in chameleon gravity has been outlined, with identified conditions for localization and examples examined, revealing asymmetric effective potentials and volcano-like structures \cite{Zhong2024}. Furthermore, the localization of a 5-dimensional gravitino field on a thick brane with nonminimal coupling was explored. In comparison to scenarios without nonminimal coupling, the inclusion of nonminimal coupling enables the localization of both the zero mode and massive Kaluza-Klein modes of a 5-dimensional free massless gravitino field to the brane \cite{Zhou2023}.

Our research objective is to investigate how the nonminimally gravity-coupled term namely scalar field with $\xi R \Phi^2$, vector field with $\lambda R A_M A^M$, and spinor fields with $\eta \bar{\Psi} R \Psi$, on the two cases of scalar thick braneworld model proposed in \cite{Liang2009} impacts the localization of matter fields. 
The model is basically RS model but with the warp function is chosen not equal to $-k |y|$, meaning that the model is generated by a bulk scalar field. The corresponding metric becomes the background metric for matter fields and no backreaction effect due to the matter field is assumed.

The writing structure is given as follows:
Section 2 reviews the formulation of the scalar thick brane system and its solutions, with two examples involving a superpotential.
Sections 3-5 examine the localization of nonminimally coupled scalar, vector, and spinor fields, respectively, with gravity on the scalar thick brane.
Section 6 presents the conclusions.

	\section{Brief on a thick brane generated by scalar bulk}
In this section, we will review a thick brane generated by a scalar field, which aims to obtain a gravitational field solution in the form of a warp factor that relates the extra dimension to the brane.
Consider a thick brane model with 5-dimensional scalar bulk $\phi (x^M)$. The gravitational action is given by
\begin{equation}\label{S-grav}
	S = \int d^5 x \sqrt{g} \left( -\frac{1}{4} R + \frac{1}{2} g^{MN} \partial_M \phi \partial_N \phi - V (\phi) \right) .
\end{equation}
The first term is the Hilbert-Einstein action in 5-dimensional spacetime, followed by the bulk scalar field lagrangian density with potential $V (\phi)$.
A 5-dimensional background spacetime is defined by the following line element with $y$ extra-coordinate,
\begin{equation}\label{metrik-y}
	ds^2 = e^{2A(y)} \tilde{g}_{\mu \nu} dx^\mu dx^\nu - dy^2 ,
\end{equation}
where $ A(y) $ is the warp function of the warp factor and $ \tilde{g}_{\mu \nu} $ is the flat brane's metric tensor. 
The $M,N$ indices represent 5-dimensional bulk coordinates, and $\mu, \nu$ indices represent the 4-dimensional brane coordinates. 
The scalar field is supposed to be a function of extra-coordinate only, $\phi = \phi (y)$. Therefore,
by varying the action with respect to scalar field and metric tensor, the bulk scalar field and Einstein's equations are obtained, respectively,
\begin{equation}
	\phi'' + 4A' \phi' - \frac{d V}{d \phi} = 0 ,
\end{equation}
\begin{equation}\label{Einstein-skalar}
	A'' = - \frac{1}{3} \phi'^2 , \quad A'^2 = \frac{\phi'^2}{12} - \frac{V}{6} ,
\end{equation}
where prime denotes a derivative with respect to the extra coordinate.
To make it easier to get field solutions, the first-order formalism is used \cite{DeWolfe2000}. The scalar field potential $ V(\phi) $ is defined through a superpotential function, $ W(\phi) $, as follows
\begin{equation}\label{W}
	V(\phi) = \frac{1}{8} \left( \frac{dW (\phi)}{d \phi} \right)^2 - \frac{2}{3} W^2(\phi),
\end{equation}
where $ W(\phi) $ as an arbitrary function of $ \phi $. 
From the Einstein's equations (\ref{Einstein-skalar}) and the given potential (\ref{W}), two first-order differential equations are obtained,
\begin{align}
	\phi' & = \pm  \frac{1}{2} \frac{dW(\phi)}{d \phi} , \label{6} \\
	A'(y) & = \pm \frac{1}{3} W(\phi(y)). \label{7}
\end{align} 
If an explicit superpotential $ W(\phi) $ is chosen in (\ref{6}), then the solution of scalar field $ \phi (y) $ and the warp function $A(y)$ (\ref{7}) are obtained.
As in Ref \cite{Liang2009}, we consider two thick brane models defined from a given superpotential.
The chosen superpotentials are functions that give rise to smooth warp factors so that the corresponding models are of thick type. We consider two models defined by $W = 3a \sinh (b \phi) $ and $W= 3 a \sin (b \phi)$ so that we have thick brane models characterized by smooth (instead of sharp) warp factors. And of course, the two superpotentials are chosen also based on the consideration that the corresponding warp factors can provide solutions to hierarchy problems.

\subsection{Model 1: $ W(\phi) = 3a \sinh (b \phi) $}
Consider a superpotential $ W(\phi) = 3 a \sinh (b \phi) $, where $ a $ and $ b $ are constant parameters. Using equations (\ref{6})-(\ref{7}), the warp function and bulk scalar solutions in $ y $-coordinate are obtained \cite{Liang2009,Bazeia2009}
\begin{align}
	A(y) & = - \frac{1}{3b^2} \ln \left[\sec^2 \left(\frac{3}{2} ab^2 y \right) \right], \label{2.8} \\ 
	\phi (y) & = \sqrt{3} \ln \left[ \sec \left(\frac{ay}{2}\right) + \tan \left(\frac{ay}{2}\right) \right] .
\end{align}
By performing the extra-coordinate transformation $ dz = e^{-A(y)} dy $, and the parameter $b^2$ is set to $\frac{1}{3}$, the relation between the $y$ and $z$ coordinates is
\begin{equation}\label{2.10}
	z(y) = \int \sec^2 \left(\frac{1}{2} ay\right) dy = \frac{2}{a} \tan \left(\frac{ay}{2}\right).
\end{equation}
The choice of parameter $b$ is related to the choice of $z$ coordinate that allows the inversion $y(z)$ to be obtained.
Taking the inverse (\ref{2.10}) and substituting $y$ into solutions $A(y)$ and $\phi(y)$, the warp function and the bulk scalar become
\begin{align}
	A(z) & = - \ln \left(1 + \frac{a^2 z^2}{4} \right), \label{warp-1}\\ 
	\phi (z) & = \sqrt{3} \ln \left( \sqrt{1 + \frac{a^2 z^2}{4}} + \frac{az}{2}  \right) . \label{bulk-1}
\end{align}
The warp function (\ref{warp-1}) gives a smooth and symmetical warp factor decending from $z=0$ to $z = \infty$. Note that the extra-coordinate transformation is taken so that we obtain simpler warp function (\ref{warp-1}) as compared to (\ref{2.8}). In addition, the extra-coordinate transformation changes the RS metric (\ref{metrik-y}) into the MRS metric. We know that the MRS thin braneworld model has better localization properties as compared to the RS one. We hope that this also happens in thick braneworld models.

\subsection{Model 2: $ W(\phi) =  3 a \sin (b \phi) $}
The second model is characterized by a superpotential function $ W(\phi) =  3a \sin (b \phi) $ where $a$ and $b$ are constant parameters. In the same way, using the equations (\ref{6})-(\ref{7}), the extra-coordinate transformation, and setting $b^2 = \frac{2}{3}$, the superpotential gives \cite{Liang2009,Bazeia2009} 
\begin{align}
	A(z) & = - \ln \left( q \sqrt{1 + \frac{a^2 z^2}{q^2}} \right) , \label{warp-2} \\
	\phi (z) & = \sqrt{\frac{3}{2}} \arcsin \left[\tanh \left({\rm arcsinh} \left(\frac{az}{q} \right) \right)\right], \label{bulk-2}
\end{align}
where $q$ is a positive integration constant.
As in model 1, we have set the parameter $b$ so that the inverse transformation $y(z)$ can be obtained. The warp function (\ref{warp-2}) quite differs from (\ref{warp-1}). Only for $q = 2$, both warp functions have a simple relationship, that is $A^{(1)} (z) = 2 A^{(2)} (z) + \ln 4$. 
In the next sections we will look for localization properties of matter fields in the two thick braneworld models.

\section{Localization of nonminimally coupled scalar field}
In this section, we will investigate the localization of the scalar field coupled nonminimally to gravity on a thick brane generated by scalar bulk. The action of nonminimally coupled scalar matter leads to the field equation, called Schr\"odinger-like equation. From the scalar field equation, we will analyze massless (zero mode) and massive fields.

\subsection{Action}
We generalized the action of the 5-dimensional scalar massless matter field $\Phi (x^M)$ considered in \cite{Liang2009}, by adding a nonminimal coupling term characterized by a coupling constant $\xi$,
\begin{equation}\label{aksi-skalar}
	S_0 = \int d^5 x \sqrt{g} \left( \frac{1}{2} g^{MN} \partial_M \Phi \partial_N \Phi + \xi R \Phi^2 \right) .
\end{equation}
The nonminimal coupling is between the scalar matter field $\Phi$ and the gravity through the curvature of spacetime $R$. Consider a braneworld metric with $z$ extra coordinate that is given in the following line element, 
\begin{equation}\label{metrik}
	ds^2 = e^{2A(z)} \left( \tilde{g}_{\mu \nu} dx^\mu dx^\nu - dz^2 \right).
\end{equation}
The extra-coordinate transformation $dy = e^{A(z)} dz$ relates the MRS metric (\ref{metrik}) with the RS metric (\ref{metrik-y}).
By this metric (\ref{metrik}), the Ricci scalar is obtained
\begin{equation}\label{Ricci}
	R = e^{-2A} \left(8 A'' + 12 A'^2\right).
\end{equation}
Using the scalar field decomposition, 
\begin{equation}\label{dekomposisi}
	\Phi (x^M) = \varphi (x^\mu) \chi (z) ,
\end{equation}
the action (\ref{aksi-skalar}) with metric (\ref{metrik}) can be rewritten as
\begin{multline}
	S_0^{(5D)} = \frac{1}{2} \int d^4 x \sqrt{\tilde{g}} \tilde{g}^{\mu \nu} \partial_\mu \varphi \partial_\nu \varphi \int_{-\infty}^\infty dz e^{3A} \chi^2 \\ + \frac{1}{2} \int d^4x \sqrt{\tilde{g}} \varphi^2  \int_{-\infty}^\infty dz \left( e^{3A} \chi'^2 + 2 \xi e^{5A} R \chi^2  \right) .
\end{multline}
It can be reduced into a 4-dimensional action
\begin{equation}\label{3.6}
	S_0^{(4D)} = \frac{1}{2} \int d^4 x \sqrt{\tilde{g}} \left( \tilde{g}^{\mu \nu} \partial_\mu \varphi \partial_\nu \varphi 
	+ m^2 \varphi^2 \right)
\end{equation}
provided that it satisfies the following normalization and mass conditions
\begin{align}
	N_0 & = \int_{-\infty}^\infty dz e^{3A} \chi^2 = 1 , \\
	m^2 & = \int_{-\infty}^\infty dz \left( e^{3A} \chi'^2 - 2 \xi e^{5A} R \chi^2 \right) < \infty . \label{3.8}
\end{align}
Note that $\varphi$ in (\ref{3.6}) describes a scalar field in a 4-dimensional spacetime of mass $m$ that interacts minimally with gravity. The appearance of mass is due to the function $\chi$ in (\ref{3.8}) in a manner similar to the mass-generation mechanism of Nambu and Jona-Lasinio \cite{Jones2013, Nambu1961}.
We define a new field $\tilde{\chi}$ to rewrite the conditions in a neatly form
\begin{equation}\label{skalar-baru}
	\tilde{\chi} (z) = e^{\frac{3}{2} A(z)} \chi (z).
\end{equation}
Then, the normalization and mass conditions become
\begin{align}
	N_0 & = \int_{-\infty}^\infty dz \tilde{\chi}^2 = 1 , \label{N0} \\
	m_0^2 & = \int_{-\infty}^\infty dz \left[ \tilde{\chi}'^2 - 3 A' \tilde{\chi}' \tilde{\chi} \right. \nonumber \\
	& \left. + \left( \frac{9}{4} A'^2 - 16 \xi A'' - 24 \xi A'^2\right) \tilde{\chi}^2 \right] < \infty. \label{m0}
\end{align}
The field $\tilde{\chi} (z)$ must satisfy these conditions in order for the scalar field $\Phi$ to be localized on the brane. 

\subsection{Field equation}
By varying the action (\ref{aksi-skalar}) with respect to the scalar field $\Phi$, the equation of motion for the 5-dimensional scalar field coupled nonminimally to gravity is obtained, 
\begin{equation}\label{3.12}
	\partial_M \left(\sqrt{g} g^{MN} \partial_N \Phi \right) - \sqrt{g} \xi R \Phi = 0.
\end{equation}
Using the decomposition (\ref{dekomposisi}), the equation (\ref{3.12}) can be rewritten into
\begin{equation}
	\frac{1}{\sqrt{\tilde{g}}} \partial_\mu \left(\sqrt{\tilde{g}} \tilde{g}^{\mu \nu} \partial_\nu \varphi \right) \chi - \left( \chi'' + 3A' \chi' \right) \varphi - \xi e^{2A} R \varphi \chi = 0 .
\end{equation}
The first term on the left reminds us the Klein-Gordon's equation in a 4-dimensional spacetimes. Thus, it is natural to consider $\varphi$ as a 4-dimensional Klein-Gordon scalar field of mass $m$, 
\begin{equation}
	\frac{1}{\sqrt{\tilde{g}}} \partial_\mu \left(\sqrt{\tilde{g}} \tilde{g}^{\mu \nu} \partial_\nu \varphi \right) + m^2 \varphi = 0.
\end{equation}
Accordingly, the "bulk component" of the scalar field fulfils the following equation
\begin{equation}
	\chi'' + 3A' \chi' + \left(m^2 + \xi e^{2A} R \right) \chi = 0. 
\end{equation}
Recalling the Ricci scalar (\ref{Ricci}), it becomes
\begin{equation}\label{field-equation}
	\chi'' + 3A' \chi' + \left[ m^2 + \xi \left(8 A'' + 12 A'^2\right) \right] \chi = 0 .
\end{equation}
From the new scalar field (\ref{skalar-baru}), the field equation (\ref{field-equation}) can be rewritten into Schr\"odinger-like equation,
\begin{equation}\label{schrodinger-skalar}
	\left[-\partial_z^2 + V_0(z)  \right] \tilde{\chi} (z) = m^2 \tilde{\chi} (z) ,
\end{equation}
where the effective potential is 
\begin{equation}\label{Potensial-skalar}
	V_0(z) = \left(\frac{3}{2} - 8 \xi \right) A'' + \left(\frac{9}{4} - 12 \xi \right) A'^2 .
\end{equation}
This effective potential, which is dependent on the warp function $A(z)$ and the coupling constant $\xi$, is primarily influenced by the warp function.
Here, we will examine  two models discussed in the previous section.

\subsection{Field localization}
\subsubsection{Model 1}\label{scalar-model-1}
\begin{figure}
	\centering
	\includegraphics[width=0.9\linewidth]{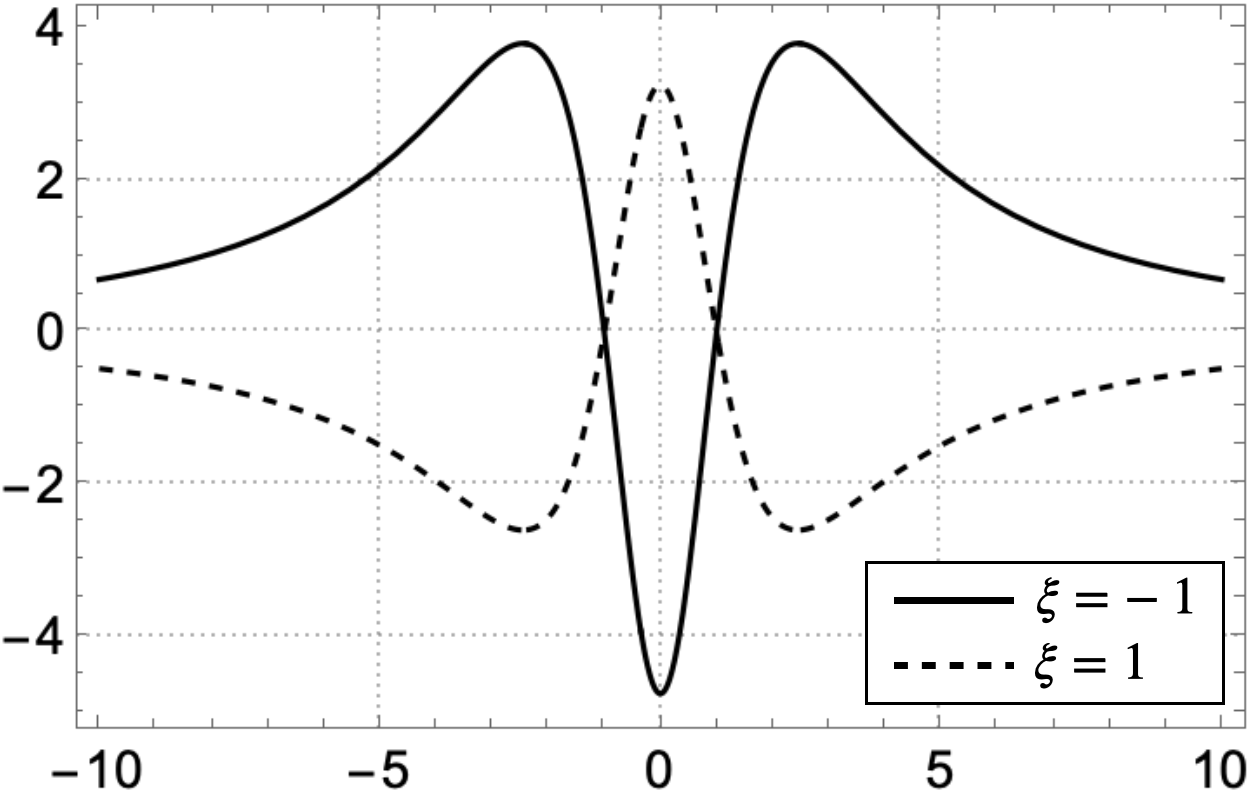}  \\
	\includegraphics[width=0.9\linewidth]{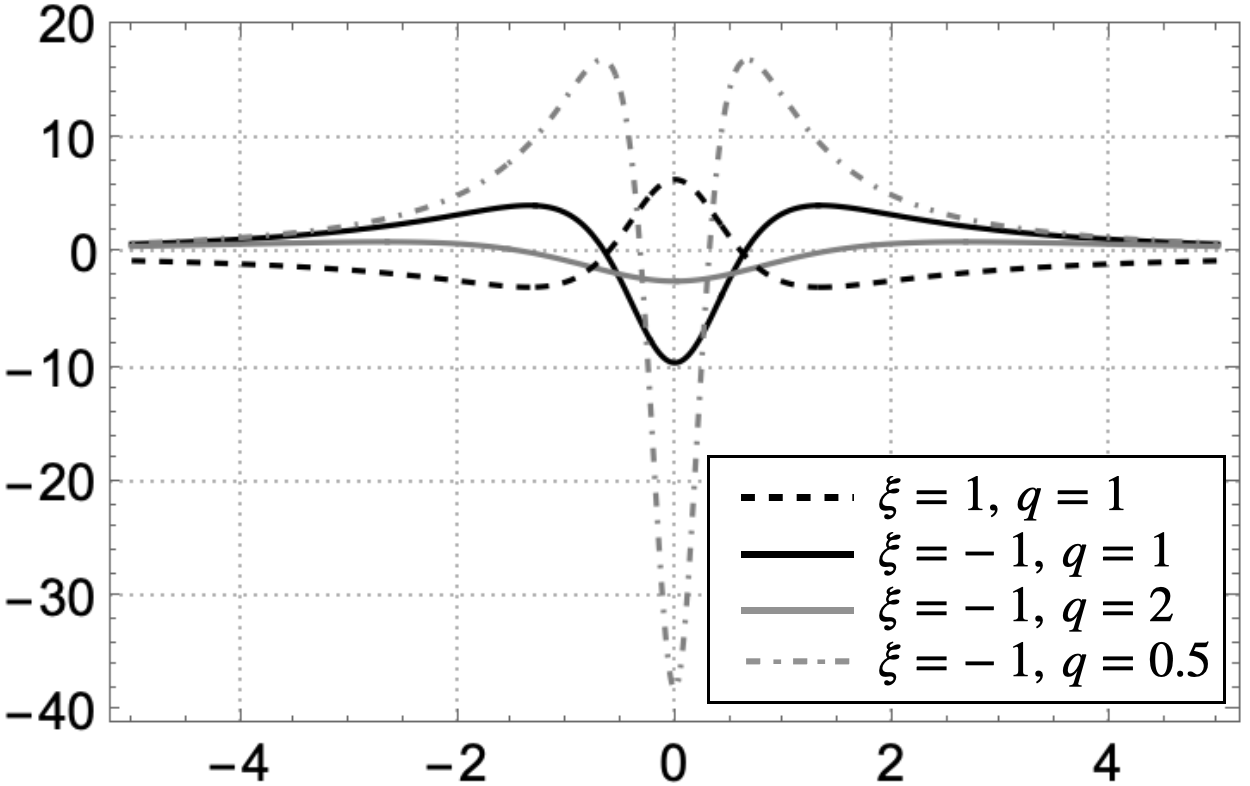}
	\caption{The shapes of the potential of the nonminimally coupled scalar field in case of Model 1 (top) and Model 2 (bottom).} 
	\label{fig:potensial-skalar-1}
\end{figure}

The warp function solution for model 1 in equation (\ref{warp-1}) leads the effective potential (\ref{Potensial-skalar}) into 
\begin{equation}\label{potensial-skalar-1}
	V_0 (z) = -\frac{4 a^2 (16 \xi -3) \left(a^2 z^2-1\right)}{\left(a^2 z^2+4\right)^2}.
\end{equation}
When the effective potential at its minima takes on negative values, the scalar field can be localized on the brane.
This potential (\ref{potensial-skalar-1}) vanishes if $a = 0$, or $z = \pm \frac{1}{a} $ where $a \neq 0$, or the coupling constant $\xi = \frac{3}{16}$. In general, the potential gives a minimum value of
\begin{equation}
	V_0^{\text{min}} = \begin{cases}
		\frac{1}{5} a^2 (3 - 16 \xi ); &  \xi > \frac{3}{16} , \\
		-\frac{1}{4} a^2 (3 - 16 \xi); &  \xi < \frac{3}{16} .
	\end{cases}
\end{equation}
The plot of the potential (\ref{potensial-skalar-1}) can be seen in Figure \ref{fig:potensial-skalar-1} (top) for the cases of coupling constants $\xi = -1$ (solid line) and $\xi = 1$ (dashed line). In general, the shapes of potentials for $\xi < \frac{3}{16}$ and  $\xi > \frac{3}{16}$ are similar to the solid line and dashed line in Figure \ref{fig:potensial-skalar-1} (top) respectively.
From the potential (\ref{potensial-skalar-1}), we will seek the possibility of the scalar matter fields localizing on the brane for all values of coupling constant except $\xi = \frac{3}{16}$.
Note that for the case of $\xi = \frac{3}{16}$, the Schr\"odinger-like equation (\ref{schrodinger-skalar}) gives sinusoidal solutions that does not fulfil the normalization condition (\ref{N0}), meaning that the scalar matter field is not localized on the brane for this case.

\paragraph{Minimal coupling}
First we consider $ \xi = 0 $, the minimal coupling case. The massless solution of (\ref{schrodinger-skalar}) is 
\begin{equation}
	\tilde{\chi}_0 (z) = \sqrt{\frac{a}{3\pi}} \frac{16}{\left(a^2 z^2+4\right)^{3/2}}.
\end{equation}
This solution fulfils the normalization condition (\ref{N0}) and the mass equation (\ref{m0}) for $m^2 = 0$, so the massless scalar field is localized on the brane. This is in accordance with \cite{Liang2009}.
On the other hand, the massive scalar field ($m\neq 0$) solution cannot analytically be obtained and thus we take an asymptotic analysis for seeking localization properties for this case and for nonminimal coupling case.

\paragraph{Asymptotic analysis}
Following Refs \cite{Guo2012,Moazzen2017}, we will discuss asymptotic behavior with approximation over a certain $z$ without losing the important point regarding normalization conditions.
In the vicinity of $z = 0$, the approximate warp function (\ref{warp-1}) is
\begin{equation}
	A(z\rightarrow 0) = -\frac{1}{4} a^2 z^2,
\end{equation} 
and correspondingly, the potential (\ref{Potensial-skalar}) in the vicinity of $z = 0$ is 
\begin{equation}
	V_0 (z \rightarrow 0) = -\frac{1}{16} a^2 (16 \xi-3) \left(3 a^2 z^2-4\right).
\end{equation}
It gives the same minimum value of the potential $V_0(z)$ for $\xi < \frac{3}{16}$, as mentioned previously. The above potential is just like a simple harmonic oscillator potential and it has a well-behaved solution. 
The approximate warp function (\ref{warp-2}) at $z \rightarrow \pm \infty$ is 
\begin{equation}\label{A-1}
	A(z \rightarrow \pm \infty) = -\ln \left(\frac{1}{4} a^2 z^2\right) ,
\end{equation}
and the corresponding potential around infinities is
\begin{equation}
	V_0 (z \rightarrow \pm \infty) = 4 \left(3-16\xi\right) z^{-2} .
\end{equation}
For the massless ($m = 0$) scalar, the field equation (\ref{schrodinger-skalar}) gives a general solution of
\begin{equation}\label{3.26}
	\tilde{\chi}_0 (z \rightarrow \pm \infty)
	= c_1 z^{\frac{1}{2} (1 - \beta)} + c_2 z^{\frac{1}{2} (1 + \beta)} ,
\end{equation}
where $\beta = \sqrt{49-256 \xi}$. 
Without loss of generality, we take $\beta$ positive and $c_2=0$. Recalling the approximate method in \cite{Guo2012,Moazzen2017}, we evaluate the normalization condition by considering the oscillator harmonic solution for integration from $z=0$ to $z=1$ and the above solution for $z=1$ to infinity. With this method, the normalization condition is fulfilled. 
The mass equation can be written as
\begin{equation}\label{3.28}
	m_0^2 = \int_{-\infty}^{-\epsilon} F(\xi,z) dz + M^2 + \int_{\epsilon}^{\infty} F(\xi,z) dz = 0 ,
\end{equation}
where $\epsilon$ is a positive number and $M^2$ is equal to the integration of $F (\xi, z)$ with respect to $z$ from $-\epsilon$ to $\epsilon$. The function $F(\xi ,z)$ is the integrand of equation (\ref{m0}) which using $A$ in (\ref{A-1}) and $\tilde{\chi}_0$ in (\ref{3.26}), it becomes
\begin{equation}\label{3.29}
	F(\xi, z) = \frac{c_1^2}{4} \left(\beta ^2-14 \beta -512 \xi +49\right) z^{-\beta -1} ,
\end{equation}
for very large $|z|$. Inserting (\ref{3.29}) into (\ref{3.28}), it gives
\begin{equation}
	M^2 = \frac{c_1^2}{2 \beta} \left(384 \xi +7 \left(\beta-7\right)\right) 
	\frac{(-\epsilon )^{\beta}-\epsilon ^{\beta}}{\left(-\epsilon ^2\right)^{\beta}}.
\end{equation}
The real value of $M^2$ is guaranteed for $\xi < \frac{49}{256}$. Thus, as $\frac{3}{16} < \frac{49}{256}$ we conclude that massless scalar fields nonminimally coupled to gravity is localized on the brane for the case $\xi < \frac{3}{16}$.
This in accordance with our previous conclusion. Inaccuracy of the approximate analysis is in the range of $\frac{3}{16} <\xi< \frac{49}{256}$. In this range, the field is localized according to the approximate analysis but it is not according to our previous analysis.

Another mode, the massive scalar field equation (\ref{schrodinger-skalar}) gives an approximate solution
\begin{equation}\label{42}
	\tilde{\chi} (z \rightarrow \pm \infty) = \sqrt{z} \left[ c_1 J_{\frac{1}{2} \beta }(m z) + c_2 Y_{\frac{1}{2} \beta}(m z)\right] ,
\end{equation}
where $J_n (z)$ is the Bessel function of the first kind, and $Y_n(z)$ is the Bessel function of the second kind. Choosing $c_2 = 0$, the massive scalar field (\ref{42}) fulfills square integrability 
for positive 4-dimensional scalar mass, $m > 0$.
It satisfies the normalization condition. Therefore, the approximate massive scalar field coupled nonminimally to gravity are localized on the scalar thick brane. 

If we consider $\xi = 0$, the Sch\"rodinger-like (\ref{schrodinger-skalar}) gives an approximate solution for the massive scalar field,
\begin{multline}
	\tilde{\chi} (z\rightarrow \pm \infty) = \frac{\sqrt{2/\pi }}{m^{\frac{7}{2}} z^3} \\ \times \left[ \left(m z \left(c_1 \left(m^2 z^2-15\right)+6 c_2 m z\right)-15 c_2\right) \cos (m z) \right. \\
	\left. + \left(c_1 \left(15-6 m^2 z^2\right)+c_2 m z \left(m^2 z^2-15\right)\right) \sin (m z)\right] ,
\end{multline}
where $c_1$ and $c_2$ are constants. This solution oscillates infinitely. It does not meet localization conditions. Thus, the massive mode minimally coupled scalar field is not localized on the brane.

\subsubsection{Model 2}
The warp function (\ref{warp-2}) for the second model gives the effective scalar potential
\begin{equation}\label{potensial-skalar-2}
	V_0 (z) = \frac{(16 \xi -3) \left(2 a^2 q^2-5 a^4 z^2\right)}{4 \left(a^2 z^2+q^2\right)^2}.
\end{equation}
Localization of the scalar field on the brane is possible when the effective potential reaches negative values at its minima. The plot of the potential (\ref{potensial-skalar-2}) is given in Figure \ref{fig:potensial-skalar-1} (bottom). Both models have similar potential shapes. From the warp function, one of the significant differentiating factors is the parameter $q$ which determines the high and low volcano potential. A smaller $q$ parameter gives a lower potential minimum, and vice versa.
There are several possible minimum values for the potential, which depend on the constant parameters $\xi$ and $q$, ($a \neq 0$)
\begin{equation}
	V_0^\text{min} = 
	\begin{cases}
		-\dfrac{25}{112 q^2}  \left(16 a^2 \xi -3 a^2\right); & \xi > \frac{3}{16}, \ q \neq 0 , \\
		\dfrac{1}{2 q^2} (16 a^2 \xi -3 a^2) ; &  \xi < \frac{3}{16}, \ q \neq 0 , \\
		-\infty ; &  \xi > \frac{3}{16} , \ q= 0 .
	\end{cases}
\end{equation}
Otherwise, the effective potential vanishes.
Therefore, the potential (\ref{potensial-skalar-2}) has possibility to localize the nonminimally coupled scalar matter.

\paragraph{Minimal coupling}
The solution of equation (\ref{schrodinger-skalar}) for minimal coupling $\xi = 0$ and massless mode is  
\begin{equation}
	\tilde{\chi}_0 (z) = \sqrt{\frac{a}{2}} \frac{q}{\left(a^2 z^2+q^2\right)^{3/4}}.
\end{equation}
It can be proved that this solution can be normalized and satisfies the mass equation for $m^2 = 0$, so that the massless scalar field coupled minimally with gravity can be localized. However, the solution of massive mode cannot be obtained analytically \cite{Liang2009}.
As in the first model, we take an asymptotic analysis for seeking localization properties for massive mode and both modes in nonminimal coupling.

\paragraph{Asymptotic analysis}
In the same way as in the previous model, the asymptotic behavior is discussed.
The warp function (\ref{warp-2}) is asymptotic to the following function as $z$ becomes very large,
\begin{equation}\label{A-2}
	A(z\rightarrow \pm \infty) = -\ln \left(az\right) ,
\end{equation}
with the derivatives
\begin{equation}
	A'(z\rightarrow \pm \infty) = - \frac{1}{z}, \quad A''(z\rightarrow \pm \infty) = \frac{1}{z^2} .
\end{equation}
The warp function approximation in this model is similar to the warp function approximation in the sine-Gordon brane model \cite{Cruz2016}. The localization conditions for a nonminimally coupled scalar field with gravity have been discussed in \cite{Moazzen2017}. Here, the scalar action is considered without the bulk scalar field mass term, $m_s = 0$. Since the normalization equation is independent of any function other than the scalar function $\tilde{\chi}_0$, the localization result should be similar to \cite{Moazzen2017}.
The zero mode scalar field is localized on the brane.

The massive mode of the nonminimally coupled scalar field leads to the following approximate solution
\begin{equation}
	\tilde{\chi} (z \rightarrow \pm \infty) = \sqrt{z} \left[ c_1 J_{\frac{1}{2} \zeta}(m z)+c_2 Y_{\frac{1}{2} \zeta}(m z)\right]
\end{equation}
where the constant parameter $\zeta = 4 \sqrt{1-5 \xi}$. 
This solution is similar to (\ref{42}), the solution in model 1, with the only difference is in the constant parameter $\zeta$. Hence, as far as the asymptotic approximate method is concerned, the nonminimally coupled massive scalar field in this model is localized on the brane.
For $\xi = 0$, the solution obtained is similar to model 1, which oscillates infinitely. Therefore, in model 2 the massive scalar field minimally coupled to gravity is not localized.

\section{Localization of nonminimally coupled vector field}
\subsection{Action}
Consider an action of vector field $A^M$ that is nonminimally coupled to gravity,
\begin{multline}
	S_1 = \int d^5x \sqrt{g} \left(-\frac{1}{4}  g^{MN} g^{RS} F_{MR} F_{NS} \right. \\ \left. + \lambda R g^{MN} A_M A_N \right) ,
\end{multline}
where $ \lambda $ is a coupling constant and $ F_{MN} = \partial_M A_N - \partial_N A_M $ is the field strength in 5-dimensional spacetime. Decomposing the vector field according to 
\begin{equation}\label{dekomposisi-vektor}
	A_M (x^M) = \left(A_\mu (x^M), A_z \right) = \left(a_\mu (x^\mu) \alpha(z), A_z \right)
\end{equation}
the non-zero components of field strength are
\begin{align*}
	F_{\mu \rho} & = \alpha \partial_\mu a_\rho - \alpha \partial_\rho a_\mu = \alpha f_{\mu \rho} , \\
	F_{\mu z} & = - a_\mu \partial_z \alpha = - F_{z \mu},
\end{align*}
where we take $A_z$, the extra coordinate component of the field, is a constant as in \cite{Jones2013}, and $f_{\mu \rho}$ is the field strength on the 4-dimensional brane. Now, the action $ S_1 $ can be rewritten as 
\begin{multline}\label{4.3}
	S_1 = -\frac{1}{4} \int d^4x \sqrt{\tilde{g}} \tilde{g}^{\mu \nu} \tilde{g}^{\rho\sigma} f_{\mu \rho} f_{\nu \sigma} \int_{-\infty}^\infty dz e^A \alpha^2 \\ 
	+ \frac{1}{2} \int d^4x \sqrt{\tilde{g}} \tilde{g}^{\mu \nu} a_\mu a_\nu \int_{-\infty}^\infty dz e^A (\partial_z \alpha)^2 \\
	+ \lambda \int d^4 x \sqrt{\tilde{g}} \tilde{g}^{\mu \nu} a_\mu a_\nu \int_{-\infty}^\infty dz e^{3A} \alpha^2 R \\
	- \lambda \int d^4x \sqrt{\tilde{g}} \int_{-\infty}^\infty dz e^{3A} R A_z^2 .
\end{multline}
In the above expressions, the first term is the kinetic term while the second and the third terms constitute the mass terms. The last term does not contain 4-dimensional vector fields and thus it does not contribute to the dynamics of the field. In addition, this term diverges due to the 4-dimensional integral part. So we can take this term to vanish by setting $A_z=0$. This setting is similar to the temporal gauge condition $a_0=0$ in the standard 4-dimensional gauge theory. In fact, we should do such setting because otherwise we would encounter imaginary mass term in the 5-dimensional field equation \cite{Liang2009}.
Removing $A_z$ also does not disturb the gauge invariance of the action within the effective 4-dimensional theory \cite{Davoudiasl2000}. Therefore, the 5-dimensional action can be reduced into an effective 4-dimensional action of vector field by fulfilling the localization conditions expressed by the normalization and mass equations,
\begin{align}
	N_1 & = \int_{-\infty}^\infty dz e^A \alpha^2 = 1 , \label{4.4} \\
	m_1^2 & = \int_{-\infty}^\infty dz \left(e^A \alpha'^2 + 2 \lambda e^{3A} \alpha^2 R \right) \nonumber\\
	& = \int_{-\infty}^\infty dz  e^A \left[ \alpha'^2 + 2 \lambda \alpha^2 \left(8A'' + 12 A'^2 \right) \right] < \infty . \label{4.5}
\end{align}

The above expressions can be rewritten in terms of a new field $ \tilde{\alpha} (z) = \alpha (z) e^{\frac{1}{2} A(z)} $, as the following
\begin{align}
	N_1  & = \int_{-\infty}^\infty dz \tilde{\alpha}^2 = 1 , \label{4.6} \\
	m_1^2 & = \int_{-\infty}^\infty dz \left[ \tilde{\alpha}'^2 - A' \tilde{\alpha}' \tilde{\alpha} \right. \nonumber \\
	& \left. + \tilde{ \alpha}^2 \left(16 \lambda A'' + \left( 24 \lambda + \frac{1}{4} \right) A'^2 \right) \right] < \infty . \label{4.7} 
\end{align}
Next, we will analyze whether $\tilde{\alpha}$ fulfills conditions (\ref{4.6})-(\ref{4.7}), namely whether the corresponding vector fields are localized on the brane.

\subsection{Field equation}
From the action $S_1$, the nonminimally coupled vector field equation is obtained by varying it with respect to the gauge field $A_R$, that is
\begin{equation}\label{persamaan-medan-vektor-5D}
	\frac{1}{\sqrt{g}} \partial_M \left(\sqrt{g} g^{MN} g^{RS} F_{NS} \right) + 2 \lambda R A^R = 0.
\end{equation}
Recalling that the field strength $f_{\mu \nu}$ in 4-dimensional spacetime fulfills the Proca equation 
\begin{equation}
	\frac{1}{\sqrt{\tilde{g}}} \partial_\mu \left(\sqrt{\tilde{g}} \tilde{g}^{\mu \nu} \tilde{g}^{\rho\sigma} f_{\nu \sigma} \right) + m^2 a^\rho  = 0 ,
\end{equation}
the field equation (\ref{persamaan-medan-vektor-5D}) is equivalent to
\begin{equation}\label{4.11}
	\alpha'' + A' \alpha' + m^2 \alpha - 2 \lambda e^{2A} R \alpha = 0
\end{equation}
and
\begin{equation}\label{4.11-}
	\partial_\mu a^\mu \alpha' + 2 \lambda R e^{2A} A^z = 0.
\end{equation}
Equation (\ref{4.11-}) shows that setting $A_z = 0$ is equivalent to choosing Lorentz gauge condition in 4-dimensional spacetime, $\partial_\mu a^\mu = 0$.

The field equation (\ref{4.11}) can be written as a Schr\"odinger-like equation for $\tilde{\alpha}$,
\begin{equation}\label{schrodinger-vektor}
	\left[ -\partial_z^2 + V_1 (z) \right] \tilde{\alpha} (z) = m^2 \tilde{\alpha} (z)
\end{equation}
where $V_1$ is the effective potential
\begin{equation}\label{Potensial-vektor}
	V_1(z) = \frac{1}{2} A'' + \frac{1}{4} A'^2 + 2 \lambda \left(8A'' + 12 A'^2\right) .
\end{equation}
In the next subsection we will analyze the localization of the vector field in the two considered models.

\subsection{Field localization}
\subsubsection{Model 1}
\begin{figure}
	\centering
	\includegraphics[width=0.9\linewidth]{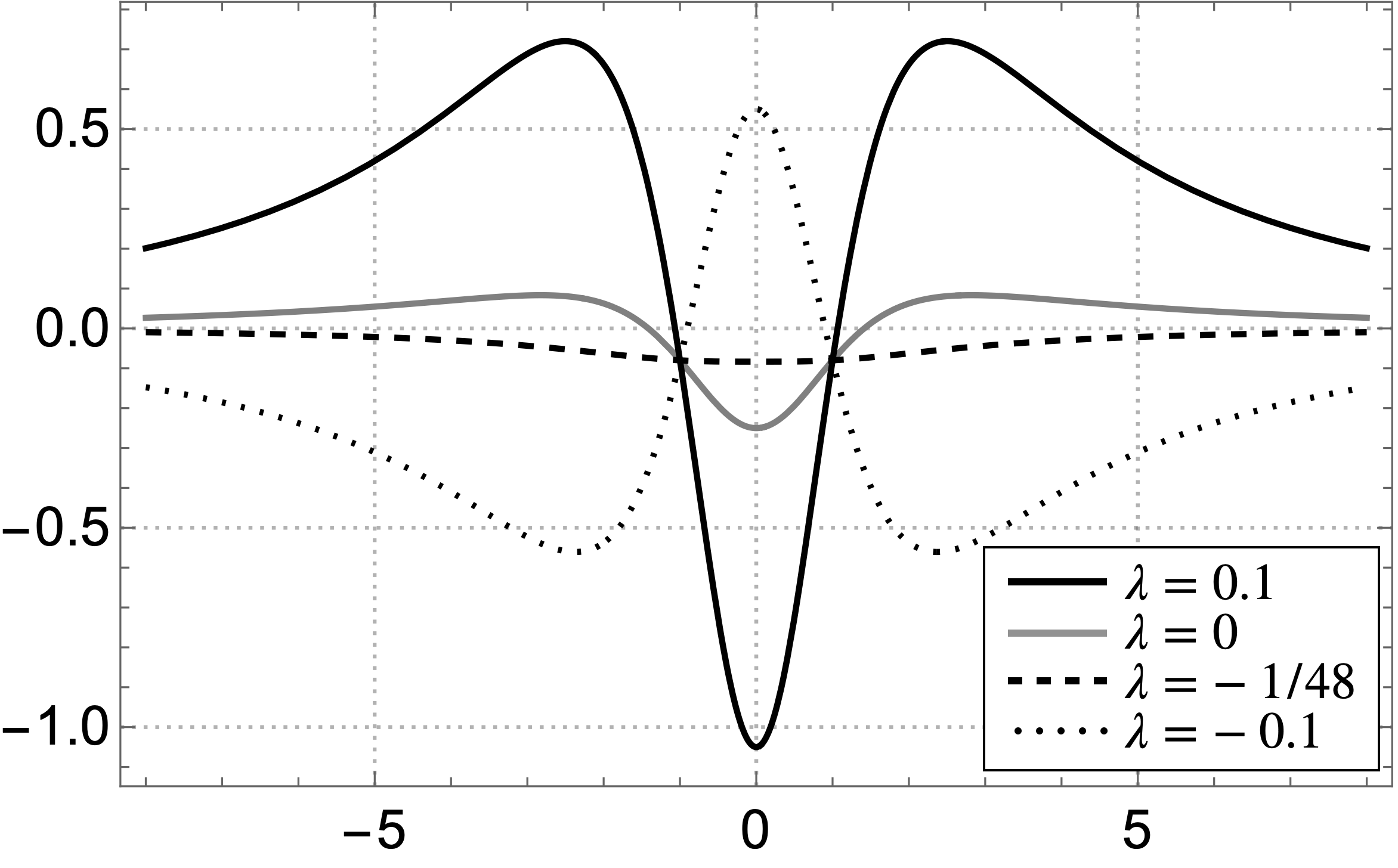} \\ 
	\includegraphics[width=0.9\linewidth]{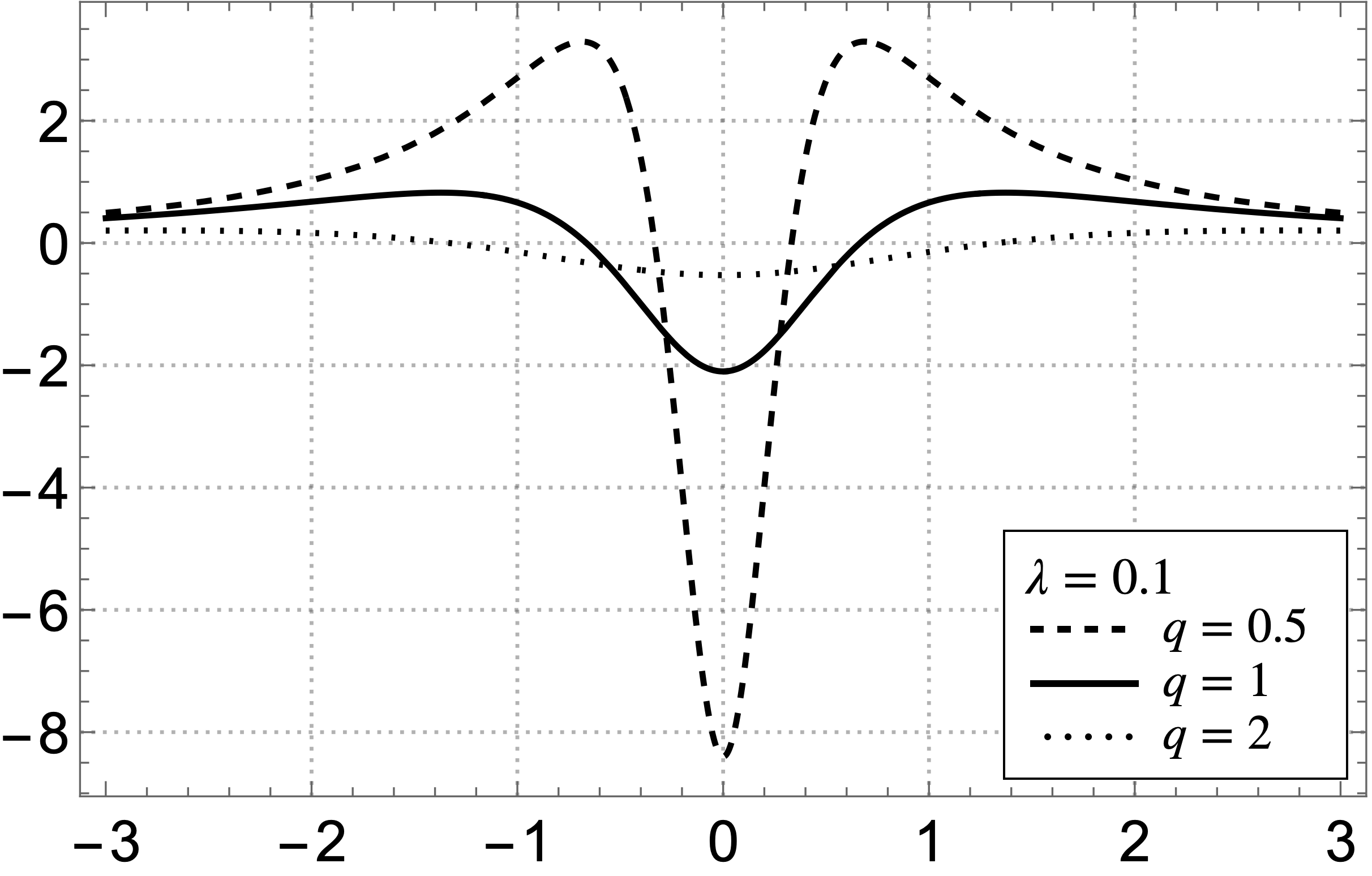}
	\caption{The shapes of the potential in the nonminimally coupled vector field: (top) Model 1 (\ref{potensial-vektor-1}) for some values of the coupling constant $\lambda$; (bottom) Model 2 (\ref{potensial-vektor-2}) for some values of the parameter $ q $, where the coupling constant $\lambda = 0.1$. The parameter $a = 1$ for both models.}
	\label{fig:potensial-vektor-1}
\end{figure}

The warp function (\ref{warp-1}) of model 1 leads the potential (\ref{Potensial-vektor}) into 
\begin{equation}\label{potensial-vektor-1}
	V_1(z) = \frac{2 a^2 \left(a^2 (64 \lambda +1) z^2-64 \lambda -2\right)}{\left(a^2 z^2+4\right)^2} .
\end{equation}
The plot of this potential (\ref{potensial-vektor-1}) is given in Figure \ref{fig:potensial-vektor-1} (top) for some values of parameters $a$ and $\lambda$. 
The parameter $a$ determines the width of the effective potential. For any real values of parameter $a$ other than zero, the potential has negative minima,
\begin{equation}
	V_1^{\text{min}} = \begin{cases}
		\frac{1}{4} \left(-32 a^2 \lambda -a^2\right)  & \text{ for } \lambda > -\frac{1}{48} , \\
		\frac{4096 a^2 \lambda ^2+128 a^2 \lambda +a^2}{4 (160 \lambda +3)} & \text{ for } \lambda \leq -\frac{1}{48} .
	\end{cases}
\end{equation}
Note that as shown in Figure \ref{fig:potensial-vektor-1}, the parameter $\lambda$ influences the shape of the effective potential.

\paragraph{Minimal coupling}
For the case of minimal coupling and massless field, equation (\ref{schrodinger-vektor}) gives
\begin{equation}\label{4.15}
	\tilde{\alpha}_0 (z) = \sqrt{\frac{2a/\pi}{a^2 z^2+4}} .
\end{equation}
This solution is not converging but it is square integrable, $\int_{-\infty}^\infty \tilde{\alpha}_0^2 (z) dz = 1$. It also fulfills the mass condition $m_1^2 = 0$. So the zero-mode vector field, the Maxwell field, interacting minimally to gravity is localized on the brane. This is a correction to Ref [11] in which it was stated that due to non-convergence of (\ref{4.15}) the corresponding field is not localized. 

\paragraph{Asymptotic behavior}
In the same way as in the scalar field, due to the difficulty in obtaining an exact solution analytically, the asymptotic behavior is used. From the approximate warp function $A(z \rightarrow \pm \infty)$ given in equation (\ref{A-1}), the effective vector potential becomes
\begin{equation}
	V_1(z\rightarrow \pm \infty) = \frac{128 \lambda +2}{z^2}.
\end{equation}
The field equation (\ref{schrodinger-vektor}), then gives the zero mode solution
\begin{multline}
	\tilde{\alpha}_0 (z\rightarrow \pm \infty) = \\ c_1 z^{-\frac{\sqrt{-64 \lambda -1} \left(\sqrt{-(64 \lambda +1)^2}+64 \lambda  \sqrt{-512 \lambda -9}+\sqrt{-512 \lambda -9}\right)}{2 (64 \lambda +1)^{3/2}}} .
\end{multline}
This solution is square integrable in the $z$ coordinate from 1 to infinity, where $\sqrt{512 \lambda + 9} \equiv \delta \neq -2 $.
Therefore, the massless $\tilde{\alpha}_0$ is trapped in the region between the two turning points and the corresponding massless vector field is localized on the brane.

The massive vector field equation (\ref{schrodinger-vektor}) gives  
\begin{equation}
	\tilde{\alpha} (z \rightarrow \pm \infty) = \sqrt{z} \left[ c_1 J_{\frac{1}{2} \delta }(m z) + c_2 Y_{\frac{1}{2} \delta}(m z)\right] .
\end{equation}
This approximate solution is similar to the case of a massive nonminimally coupled scalar field (\ref{42}). Therefore, by following the asymptotic behavior method, we obtain that a massive vector field nonminimally coupled to gravity is localized on the brane.
Meanwhile, if we consider $\lambda = 0$, the massive mode minimally coupled vector field equation (\ref{schrodinger-vektor}) gives an oscillate solution,
\begin{multline}
	\tilde{\alpha} (z \rightarrow \pm \infty) = -\frac{\sqrt{2/(\pi m^3)}}{z} \left[ (c_1 m z+c_2) \cos (m z) \right. \\ \left. + (c_2 m z-c_1) \sin (m z) \right] .
\end{multline}
Such a solution does not guarantee the fulfillment of the condition (\ref{4.6}). Thus, the massive mode minimally coupled vector fields is not localized on the brane.

\subsubsection{Model 2}
Consider the warp function (\ref{warp-2}). It gives the effective scalar potential
\begin{equation}\label{potensial-vektor-2}
	V_1(z) = 
	\frac{a^4 (160 \lambda +3) z^2-2 a^2 (32 \lambda +1) q^2}{4 \left(a^2 z^2+q^2\right)^2} .
\end{equation}
This potential has similar shape with that in model 1. In model 2 in Figure \ref{fig:potensial-vektor-1} (bottom), we show the influence of the parameter $q$ in the potential. For a larger value of $q$, the potential will move towards zero. For non-zero constant $a$, and coupling constant $ \lambda > -\frac{7 q^2}{352} (q \neq 0)$, the above potential gives a minimum value
\begin{equation}
	V_1 = \frac{-32 a^2 \lambda -a^2 q^2}{2 q^4}. 
\end{equation}
This allows for the field localization.

\paragraph{Minimal coupling}
Now we go to the remaining cases for model 2: massless mode for minimal coupling (zero coupling constant) and for positive coupling constant.
First, we consider the coupling constant $\lambda = 0$, the zero mode solution of (\ref{schrodinger-vektor}) is
\begin{equation}
	\tilde{\alpha}_0 (z) = \frac{c_1}{\left(a^2 z^2+q^2 \right)^{1/4}} . 
\end{equation}
This solution could not be normalized on $(-\infty, \infty)$. Thus, the massless vector field is not localized on the brane for model 2 with minimal coupling.

\paragraph{Asymptotic behavior}
In the same way as in model 1, we should consider asymptotic behavior for the case of positive coupling constant and zero mode. 
The approximate warp function (\ref{A-2}) gives the effective potential
\begin{equation}
	V_1(z\rightarrow \pm \infty) = \frac{160 \lambda +3}{4 z^2}.
\end{equation}
The corresponding Schr\"odinger-like equation with zero mass has the solution of
\begin{equation}
	\tilde{\alpha}_0 (z\rightarrow \pm \infty) = c_1 z^{\frac{1}{2}-\frac{\sqrt{-160 \lambda -3} \sqrt{-40 \lambda -1}}{\sqrt{160 \lambda +3}}}.
\end{equation}
This solution satisfies square integrability in the $z$-coordinate limit from 1 to infinity.
Accordingly, as in model 1, massless vector fields are localized on the brane for nonminimal model 2 with positive coupling constants.
The massive vector field equation (\ref{schrodinger-vektor}) gives
\begin{equation}
	\tilde{ \alpha} (z\rightarrow \pm \infty) = \sqrt{z} \left(c_1 J_{\varrho}(m z)+c_2 Y_{\varrho}(m z)\right),
\end{equation}
where $ \varrho = \sqrt{40 \lambda +1}$.
Choosing $c_2 = 0 $, this solution fulfills square integrability.
For $\lambda = 0$, the massive vector field behaves similarly to model 1, so that the minimally coupled massive vector field is not localized on the brane.

\section{Localization of nonminimally coupled spinor field}
In this section, we will examine the localization of the spinor field coupled nonminimally to gravity in the two thick brane models.
As in \cite{Wulandari2019, Liang2009, Guerrero2019}, we consider spinor fields coupled nonminimally in the form of a scalar-fermion coupling $\bar{\Psi} R \Psi$.

\subsection{Action}
An action of nonminimally coupled 5-dimensional spinor $\Psi (x^M)$ with gravity is given,
\begin{equation}\label{aksi-1/2}
	S_\frac{1}{2} = \int d^5 x \sqrt{g} \left(\bar{\Psi} i \Gamma^M D_M \Psi - \eta \bar{\Psi} R \Psi \right) ,
\end{equation}
with coupling constant $ \eta $ and Ricci scalar $ R $.
Since we define the field in a curved space, the covariant derivative and non-zero spin connection $\omega_M $ are given as \cite{Oda2000, Liang2009}
\begin{equation}
	D_M = \partial_M + \omega_M ; \quad \omega_M = \omega_\mu = -\frac{i}{2} A' \gamma_\mu \gamma_5 .
\end{equation}
$ \Gamma^M = \left(e^{-A} \gamma^\mu , -i e^{-A} \gamma^5 \right) $ are 5-dimensional gamma matrices in a curved space.
We can extract the 4-dimensional theory of spinor from the 5-dimensional action (\ref{aksi-1/2}). By inserting the covariant derivative and gamma matrices into the 5-dimensional action, we can express the action as
\begin{multline}\label{5.7}
	S_{\frac{1}{2}} = \int d^5 x \sqrt{\tilde{g}} \left( \Psi^\dagger e^{3A} \gamma^0 i \gamma^\mu \partial_\mu \Psi \right. \\
	+ \Psi^\dagger e^{3A} \gamma^0 \gamma^\mu \frac{1}{2} A' \gamma_\mu \gamma_5 \Psi + \Psi^\dagger e^{3A} \gamma^0 \gamma^5 \partial_z \Psi \\
	\left. - \eta \Psi^\dagger e^{4A} \gamma^0 R \Psi \right) .
\end{multline}
The 5-dimensional spinor field can be decomposed into two chiral components, corresponding to its left-handed and right-handed projections.
Each component can be separated into brane and extra dimension parts, that is
\begin{equation}\label{spinor-decomposition}
	\Psi (x^M) = \begin{pmatrix}
		\psi_L^{(1)} (x^\mu) P_L^{(1)} (z) \\
		\psi_L^{(2)} (x^\mu) P_L^{(2)} (z) \\
		\psi_R^{(1)} (x^\mu) P_R^{(1)} (z) \\
		\psi_R^{(2)} (x^\mu) P_R^{(2)} (z)
	\end{pmatrix} .
\end{equation}
Then, the action (\ref{5.7}) is rewritten into the following forms.\\ 
The first term:
\begin{align*}
	& \int d^5 x \sqrt{\tilde{g}}  e^{3A} i \\
	& \times \left[ \sum_{j=1}^{2} \left(\psi_R^{(j)*} \partial_0 \psi_R^{(j)} P_R^{(j)*} P_R^{(j)} + \psi_L^{(j)*} \partial_0 \psi_L^{(j)} P_L^{(j)*} P_L^{(j)} \right)\right. \\
	& \ + \left( \psi_R^{(1)*} \partial_1 \psi_R^{(2)} P_R^{(1)*} P_R^{(2)} + \psi_R^{(2)*} \partial_1 \psi_R^{(1)} P_R^{(2)*} P_R^{(1)} \right. \\ 
	& \left. + \psi_L^{(1)*} \partial_1 \psi_L^{(2)} P_L^{(1)*} P_L^{(2)} + \psi_L^{(2)*} \partial_1 \psi_L^{(1)} P_L^{(2)*} P_L^{(1)} \right) \\
	& + \left( \psi_R^{(1)*} (-i\partial_2) \psi_R^{(2)} P_R^{(1)*} P_R^{(2)} + \psi_R^{(2)*} i \partial_2 \psi_R^{(1)} P_R^{(2)*} P_R^{(1)} \right. \\ 
	& \ \left. + \psi_L^{(1)*} (-i\partial_2) \psi_L^{(2)} P_L^{(1)*} P_L^{(2)} + \psi_L^{(2)*} i \partial_2 \psi_L^{(1)} P_L^{(2)*} P_L^{(1)} \right) \\
	& + \left(\psi_R^{(1)*} \partial_3 \psi_R^{(1)} P_R^{(1)*} P_R^{(1)} - \psi_R^{(2)*} \partial_3 \psi_R^{(2)} P_R^{(2)*} P_R^{(2)}  \right. \\
	& \ \left. \left. + \psi_L^{(1)*} \partial_3 \psi_L^{(1)} P_L^{(1)*} P_L^{(1)} - \psi_L^{(2)*} \partial_3 \psi_L^{(2)} P_L^{(2)*} P_L^{(2)}\right) \right] .
\end{align*}
The second term:
\begin{multline*}
	\sum_{j=1}^{2} \left[ - 2 \int d^4 x \sqrt{\tilde{g}} \psi_R^{(j)*} \psi_L^{(j)} \int dz A' e^{3A} P_R^{(j)*} P_L^{(j)} \right. \\ \left. + 2 \int d^4 x \sqrt{\tilde{g}} \psi_L^{(j)*} \psi_R^{(j)} \int dz A' e^{3A} P_L^{(j)*} P_R^{(j)} \right] .
\end{multline*}
The third term:
\begin{multline*}
	\sum_{j=1}^{2} \left[ - \int d^4 x \sqrt{\tilde{g}} \psi_R^{(j)*} \psi_L^{(j)} \int dz e^{3A} P_R^{(j)*} \partial_z P_L^{(j)}  \right. \\ \left. + \int d^4 x \sqrt{\tilde{g}} \psi_L^{(j)*} \psi_R^{(j)} \int dz e^{3A} P_L^{(j)*} \partial_z P_R^{(j)} \right] .
\end{multline*}
The fourth term:
\begin{multline*}
	\sum_{j=1}^{2} \left[ - \int d^4 x \sqrt{\tilde{g}} \psi_R^{(j)*} \psi_L^{(j)} \int dz \eta e^{4A} R P_R^{(j)*} P_L^{(j)}  \right. \\ \left. - \int d^4x \sqrt{\tilde{g}} \psi_L^{(j)*} \psi_R^{(j)} \int dz \eta e^{4A} R P_L^{(j)*} P_R^{(j)} \right].
\end{multline*}
The theory of 5-dimensional spinor with nonminimal coupling can be reduced to the following 4-dimensional spinor theory with mass $m$,
\begin{equation}\label{aksi-spinor4D}
	S^{(4D)} 
	= \int d^4x \sqrt{\tilde{g}} \left(\bar{\psi} i \gamma^\mu \partial_\mu \psi - m \bar{\psi} \psi \right),
\end{equation}
by satisfying the normalization and mass equations
\begin{subequations}\label{8}
	\begin{align}
		N_1 & = \int dz e^{3A} P_R^2 = 1 , \label{5.9a} \\
		N_2 & = \int dz e^{3A} P_L^2 = 1 , \label{5.9b} \\
		m_1 & = \int dz e^{3A} P_R \partial_z P_L < \infty , \label{5.9c} \\
		m_2 & = \int dz e^{3A} P_L \partial_z P_R  < \infty , \\
		m_3 & = \int dz A' e^{3A} P_R P_L < \infty , \\
		m_4 & = \int dz \eta e^{4A} R P_L P_R < \infty ,  \label{5.9f}
	\end{align}
\end{subequations}
where for simplicity we take $P^{(1)} = P^{(2)}$.
Note that the first term in 5-dimensional action is related to the first term in (\ref{aksi-spinor4D}) by the normalization equations (\ref{5.9a})-(\ref{5.9b}), while the second-fourth terms are associated with last term in (\ref{aksi-spinor4D}) by the mass equations (\ref{5.9c})-(\ref{5.9f}).

\subsection{Field equation}
By varying the action (\ref{aksi-1/2}) with respect to $\bar{\Psi}$, the equation of motion for the nonminimally coupled 5-dimensional spinor field is
\begin{equation}
	i \Gamma^M D_M \Psi - \eta R \Psi = 0 ,
\end{equation}
which is equivalent to
\begin{equation}\label{54}
	\left(i \gamma^\mu \partial_\mu + \gamma^5 (\partial_z + 2A') - \eta e^A R \right) \Psi = 0. 
\end{equation}
Considering the decomposition (\ref{spinor-decomposition}), where $ \psi_{L,R} (x^\mu) $ obey the Dirac equation, $ i\gamma^\mu \partial_\mu \psi_{L,R} = m \psi_{R,L} $, and $ \gamma^5 \psi_L = - \psi_L $, $  \gamma^5 \psi_R = \psi_R $, the field equation (\ref{54}) becomes
\begin{subequations}\label{5.6}
	\begin{align}
		& \left( \partial_z + 2A' - \eta e^A R \right) P_{Rn} = -m_n P_{Ln}, \\
		& \left( \partial_z + 2A' + \eta e^A R \right) P_{Ln} = m_n P_{Rn}.
	\end{align}
\end{subequations}
By defining $\tilde{P}_{L,R} = e^{2A} P_{L,R}$ the above equations reduce to Schr\"odinger-like equations
\begin{equation}\label{20}
	\left[ -\partial_z^2 + V_{L,R} \right] \tilde{P}_{L,R} = m^2 \tilde{P}_{L,R}
\end{equation}
where the effective potential $V_{L,R}$ are of the forms
\begin{equation}
	V_L = \eta^2 e^{2A} R^2 - \eta e^A (R' + A' R) ,
\end{equation}
\begin{equation}
	V_R = \eta^2 e^{2A} R^2 + \eta e^A (R' + A' R) .
\end{equation}
The normalization and mass equations (\ref{8}) in terms of $\tilde{P}_{L,R}$ are
\begin{subequations}\label{24}
	\begin{align}
		N_1 & = \int dz e^{-A} \tilde{P}_R^2 = 1 , \label{5.16a1} \\
		N_2 & = \int dz e^{-A}  \tilde{P}_L^2 = 1 , \label{5.16a}\\
		m_1 & = \int dz e^{-A} \tilde{P}_R \left( \partial_z \tilde{P}_L - \frac{3}{2} A' \tilde{P}_L \right) < \infty,  \label{5.16c} \\ 
		m_2 & = \int dz e^{-A} \tilde{P}_L \left( \partial_z \tilde{P}_R - \frac{3}{2} A' \tilde{P}_R \right) < \infty , \\
		m_3 & = \int dz e^{-A} A' \tilde{P}_R \tilde{P}_L < \infty , \\
		m_4 & = \int dz \eta R \tilde{P}_L \tilde{P}_R < \infty. \label{5.16b}
	\end{align}
\end{subequations}

\subsection{Field localization}

\subsubsection{Model 1}
\begin{figure}
	\centering
	\includegraphics[width=0.95\linewidth]{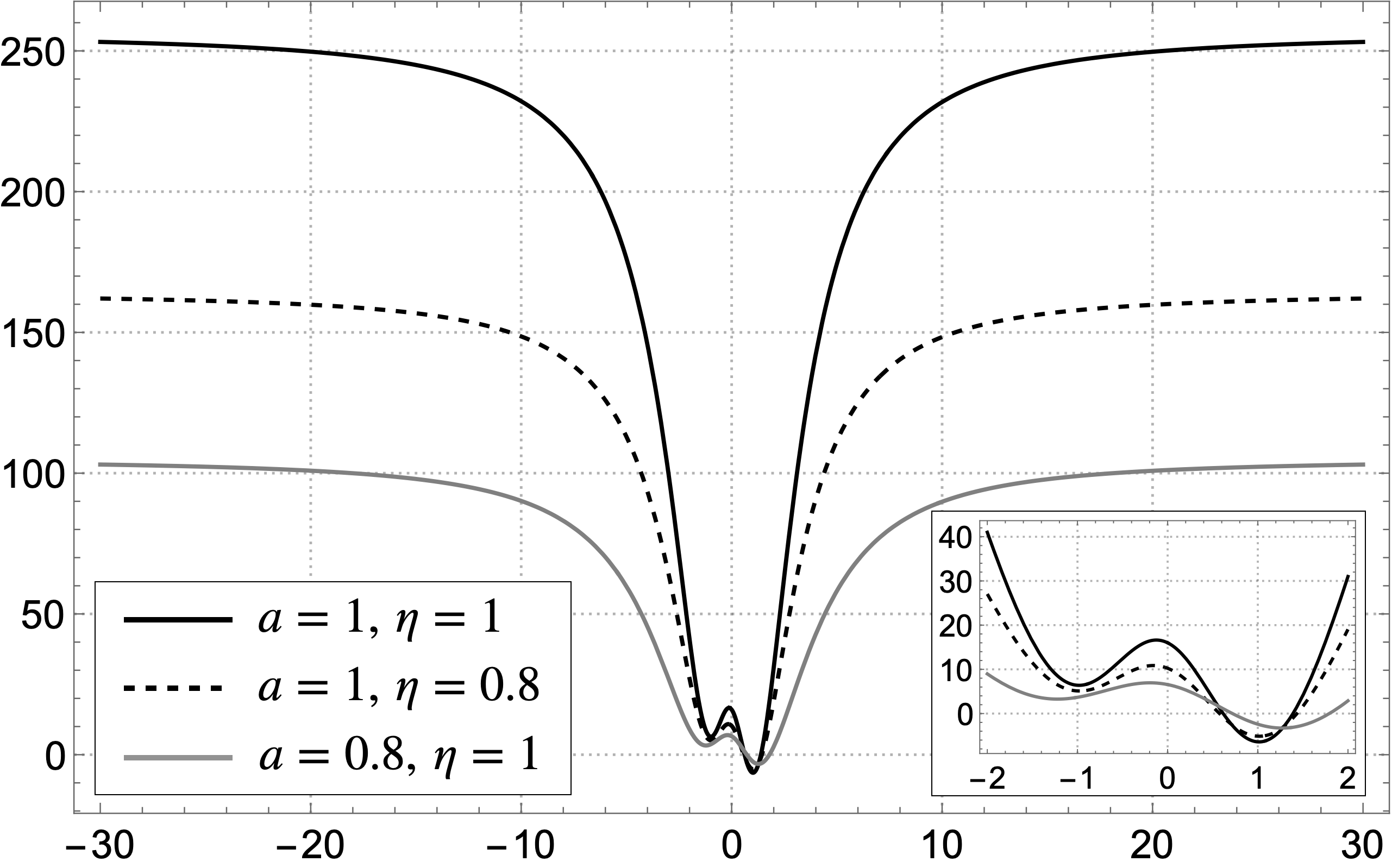}
	\caption{The shapes of potential $V_L (z)$ in Model 1 for some parameters $a$ and $\eta$.}
	\label{fig:potensial-spinorl-1}
\end{figure}

From the warp function (\ref{warp-1}), the effective potential for left-handed spinor becomes
\begin{align}
&	V_L (z) = \frac{32 a^4 \eta  \left(8 a^4 \eta  z^4-16 a^2 \eta  z^2+8 \eta -5 z\right)}{\left(a^2 z^2+4\right)^2}, \nonumber \\ 
& V_R (z) = \left. V_L (z) \right|_{\eta \rightarrow -\eta} . \label{potential-spinor-1}
\end{align}
The plot of the potential for some parameters $a$ and $\eta$ can be seen in Figure \ref{fig:potensial-spinorl-1}. The potential $V_R (z)$ has a similar shape, it is a reflection of  $V_L (z)$ about the vertical axis. 
Note that the smaller the value of $\eta$, the minimum value of $V_L$ goes to zero. For the case of minimal coupling $\eta = 0$, the potential vanishes and the solutions $\tilde{P}_{L,R}$ of (\ref{20}) are linear functions for massless mode and sinusoidal functions for massive mode. All of these functions are not square integrable and hence spinor fields interacting minimally to gravity are not localized on the brane in this model. 
Now for nonminimal coupling, equations (\ref{20}) give particular solutions for massless spinors case
\begin{equation}\label{PL0}
	\tilde{P}_{L0} (z) = c_1 \exp \left[ -8 a \eta  \left(2 a z-5 \arctan \left(\frac{a z}{2}\right)\right)\right] ,
\end{equation}
\begin{equation}\label{PR0}
	\tilde{P}_{R0} (z) = c_2 \exp \left[8 a \eta  \left(2 a z-5 \arctan \left(\frac{a z}{2}\right)\right) \right] .
\end{equation}
For any positive $\eta$, the function $\tilde{P}_{L0} (z)$ converges only as long as $z$ is positive. If we take the limit $\tilde{P}_{L0} (z)$ on $z$ toward $\pm \infty$, the function vanishes only in the positive $z$ direction.
For negative $\eta$, the function $\tilde{P}_{L0} (z)$ converges only as long as $z$ is negative. If a limit of $z \rightarrow \pm \infty$ is taken, the function vanishes only in the negative $z$ direction. Conversely for $\tilde{P}_{R0} (z)$.
Thus, using the particular solutions (\ref{PL0})-(\ref{PR0}), the massless spinor field interacting nonminimally with gravity is not localized on the brane in this model.

\paragraph{Asymptotic analysis}
In the same way as in the case of scalar and vector fields, by considering the asymptotic behavior of the warp function of model 1 in equation (\ref{A-1}), the approximate effective potential (\ref{potential-spinor-1}) is
\begin{equation}\label{VL-approx}
	V_{L,R} (z\rightarrow \pm \infty) = 256 a^4 \eta ^2.
\end{equation}
The zero mode left-handed spinor field equation gives an approximate solution
\begin{equation}
	\tilde{P}_{L0} (z \rightarrow \pm \infty) = c_1 e^{16 a^2 \eta  z} + c_2 e^{-16 a^2 \eta  z} .
\end{equation}
Since the approximate potential have the same form, the approximate solution $\tilde{P}_{R0} = \tilde{P}_{L0}$.
As $c_1$ and $c_2$ are arbitrary constants, we can take $c_1=0$. The corresponding solution fulfils the normalization equations (\ref{5.16a1})-(\ref{5.16a}).
Similar to the case of a scalar field, the mass equations in (\ref{5.16c})-(\ref{5.16b}) must satisfy the first and third terms of the following equation,
\begin{equation}
	m_\frac{1}{2}^2 = \int_{-\infty}^{-\epsilon} G_1(\xi,z) dz + M^2 + \int_{\epsilon}^{\infty} G_1(\xi,z) dz = 0.
\end{equation}
It can be proven that the mass equations (\ref{5.16c})-(\ref{5.16b}) are satisfied.

For a massive spinor, the approximate solutions are
\begin{multline}
	\tilde{P}_{L,R} (z\rightarrow \pm \infty) = c_1 e^{z \sqrt{256 a^4 \eta ^2-m^2}} \\ +c_2 e^{-z \sqrt{256 a^4 \eta ^2-m^2}} .
\end{multline}
The approximate solution of massive field also satisfies the localization conditions (\ref{24}). Thus, the spinor field can be localized for this model.

\subsubsection{Model 2}
\begin{figure}
	\centering
	\includegraphics[width=0.9\linewidth]{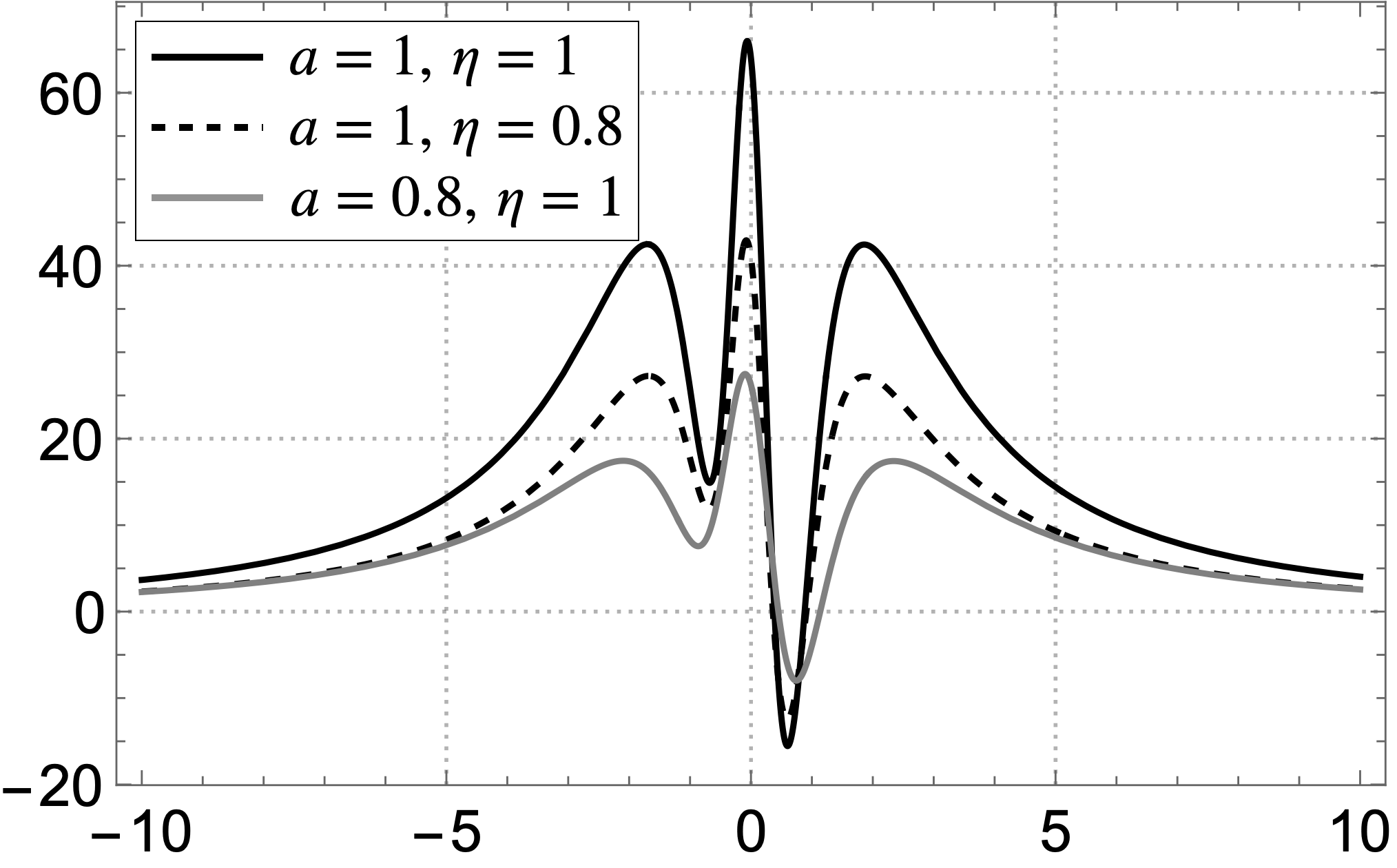} \\
	\includegraphics[width=0.9\linewidth]{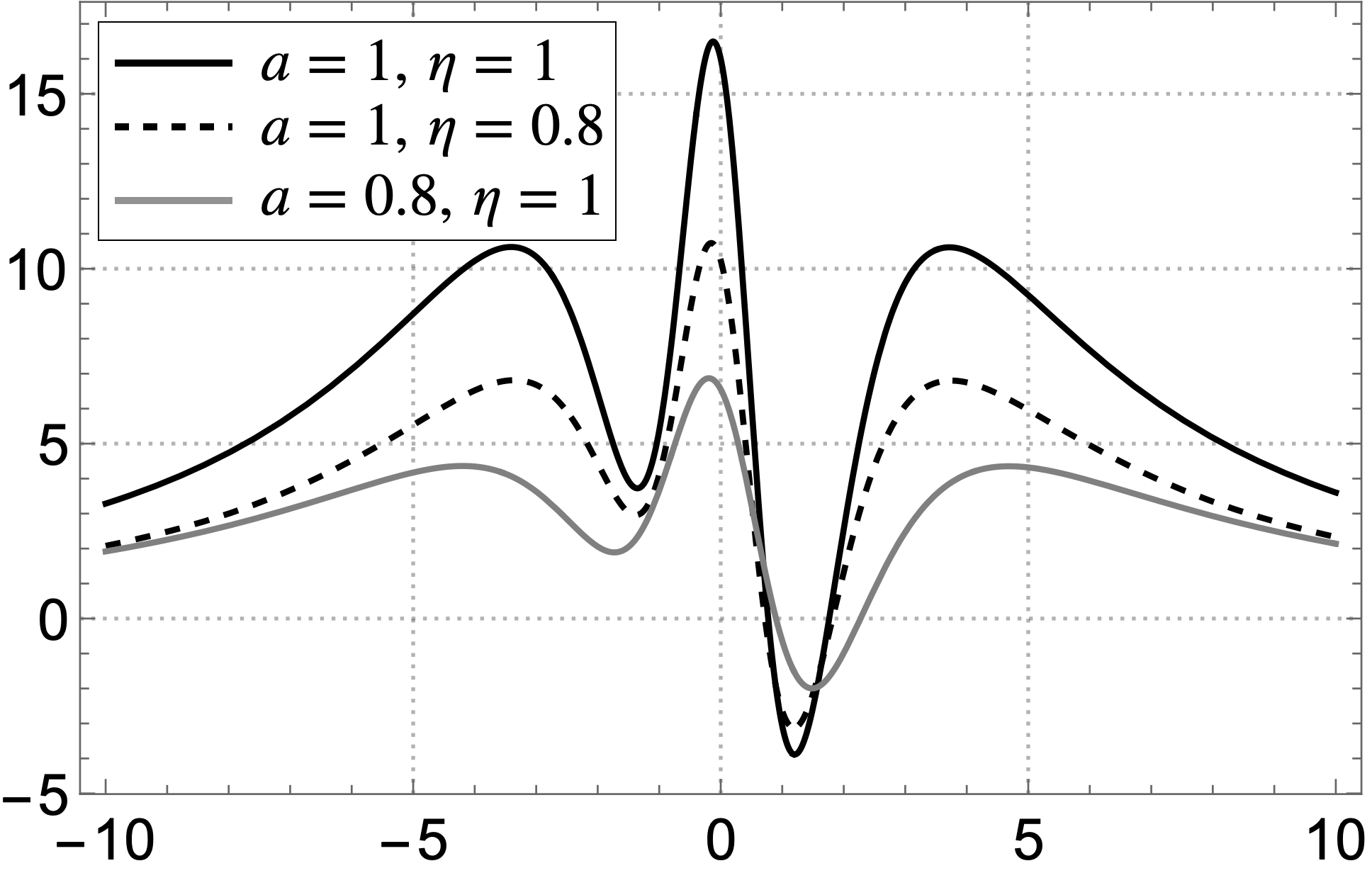}
	\caption{The potential shapes of the spinor left-handed $V_{L} (z)$ in case of Model 2. Without reducing the generality of the influence of constants, we choose $q = 1$ (top) and $q=2$ (bottom), with different constant parameters $a$ and $\eta$.}
	\label{fig:potensial-spinorl-2}
\end{figure}

Consider the warp function (\ref{warp-2}), the effective potential for left-handed spinor becomes
\begin{multline}\label{V-66}
	V_L (z) = \frac{4 a^4 \eta}{\left(a^2 z^2+q^2\right)^3} \left(100 a^4 \eta  z^4-80 a^2 \eta  q^2 z^2 \right. \\
	\left. + \left(5 a^2 q z^3 -16 q^3 z \right) \sqrt{\frac{a^2 z^2}{q^2}+1} + 16 \eta  q^4\right),
\end{multline}
The right-handed spinor's effective potential is $ V_R (z) = \left. V_L (z) \right|_{\eta \rightarrow -\eta} $.
With this potential, the solution of the equation (\ref{20}) is difficult to obtain analytically.
The shape of potential (\ref{V-66}) is given in Figure \ref{fig:potensial-spinorl-2}.

\paragraph{Asymptotic analysis}
The asymptotic behavior of warp function in (\ref{A-2}) leads the effective potential (\ref{V-66}) into an approximate form,
\begin{equation}
	V_{L} (z\rightarrow \infty) = \frac{20 \eta  \left(20 a^2 \eta + a\right)}{z^2} .
\end{equation}
The approximate solution of zero mode left-handed spinor is
\begin{equation}\label{5.19}
	\tilde{P}_{L0} (z\rightarrow \infty) = c_1 z^{\frac{1}{2}-\frac{1}{2} (40 a \eta + 1)}. 
\end{equation}
If the limit $z$ is taken to infinity, this solution vanishes. 
In order to localize the zero mode left-handed spinor on the brane, $ \tilde{P}_{L0} (z \rightarrow \infty) $ should satisfy the normalization condition.
On the other hand, the right-handed spinor needs to be examined as well because the approximate potential is not in the same form as that of the left-handed.
The approximate solution of zero mode right-handed spinor is
\begin{equation}
	\tilde{P}_{R0} (z \rightarrow \infty) = c_1 z^{\frac{1}{2} - \frac{1}{2} (40 a \eta - 1)} .
\end{equation}
Using these solutions, the normalization equations (\ref{5.16a1})-(\ref{5.16a}) are also fulfilled.
It can be proven that the mass equations in (\ref{5.16c})-(\ref{5.16b}) satisfy the first and third terms of the following equation,
\begin{equation}
	m_\frac{1}{2}^2 = \int_{-\infty}^{-\epsilon} G_2 \xi,z) dz + M^2 + \int_{\epsilon}^{\infty} G_2 (\xi,z) dz = 0.
\end{equation}
In conclusion, this model provides localized nonminimally coupled massless spinors on the scalar thick brane.

\section{Discussion and conclusions}\label{sec12}

Our investigation has focused on how Standard Model fields nonminimally coupled to gravity are confined on a thick braneworld model generated by a scalar bulk. We consider two models of scalar thick brane each derived from a different given superpotential \cite{Liang2009}. To study the localization of these fields, we have utilized a natural mechanism. The normalization condition is obtained by reducing the 5-dimensional action to a 4-dimensional action.

We consider a scalar field coupled nonminimally to gravity with a coupling constant $\xi$. The exact solutions of the field equations are difficult to obtain, and normalization analysis cannot be performed analytically. In the normalization equation, we only need to understand how the field $\tilde{\chi} (x^5)$ behaves in the extra coordinate. In this case, a square-integrable field function is required.
Without losing the point of field localization, we examine the asymptotic behavior of the warp function on $z$ towards infinity. The massless and massive modes of the nonminimally coupled scalar field are localized in both models of the scalar thick brane. In terms of the exact solution, when the coupling is set to be minimal, the scalar field is localized for the massless mode. This result confirms the minimal case as stated in \cite{Liang2009}.

In a qualitative sense, the analysis of scalar field localization can be explained as follows. Consider the volcano-shaped potential in both models. The mass term in the Schr\"odinger-like equation is analogous to energy term. The massless case is equivalent to energy $E=0$ case. In this case, the field is trapped within the potential boundaries of $-\frac{1}{a} \leq z \leq \frac{1}{a}$ where $z = \pm \frac{1}{a}$ are turning points. Meanwhile, in the case of a field with mass $m\neq0$, the field can be either trapped in a region where the brane is the mid of the region or in $d \leq  |z| < \infty$. The value of $d$ depends on the mass, the smaller the mass $m$, the closer $d$ to infinity. These two regions are separated by a potential wall of length $(d-b)$ where $b$ is the turning point in the first region. Small masses are unlikely able to tunnel the wall since due to very wide of the wall. Light mass fields are possible to localize on the brane. Likewise for other types of fields.

In the case of a nonminimally coupled vector field, it has been discovered that a vector field with a nonminimal coupling constant $\lambda$ is also localized on a scalar thick brane for both models.
In the context of a nonminimally coupled spinor field, the massless mode is not localized on the brane and the massive is localized using the asymptotic analysis in model 1. In model 2, the massless mode is localized on the brane and the massive mode is not localized.
In general, within the approximate method, nonminimally coupled fields with gravity have better localization properties than their minimally coupled counterparts.

\backmatter

\bmhead{Acknowledgments}
This work was supported by PMDSU, P3MI and PPMI-FMIPA Institut Teknologi Bandung.


\bibliography{locscalarbranes}


\begin{thebibliography}{36}
\ifx \bisbn   \undefined \def \bisbn  #1{ISBN #1}\fi
\ifx \binits  \undefined \def \binits#1{#1}\fi
\ifx \bauthor  \undefined \def \bauthor#1{#1}\fi
\ifx \batitle  \undefined \def \batitle#1{#1}\fi
\ifx \bjtitle  \undefined \def \bjtitle#1{#1}\fi
\ifx \bvolume  \undefined \def \bvolume#1{\textbf{#1}}\fi
\ifx \byear  \undefined \def \byear#1{#1}\fi
\ifx \bissue  \undefined \def \bissue#1{#1}\fi
\ifx \bfpage  \undefined \def \bfpage#1{#1}\fi
\ifx \blpage  \undefined \def \blpage #1{#1}\fi
\ifx \burl  \undefined \def \burl#1{\textsf{#1}}\fi
\ifx \doiurl  \undefined \def \doiurl#1{\url{https://doi.org/#1}}\fi
\ifx \betal  \undefined \def \betal{\textit{et al.}}\fi
\ifx \binstitute  \undefined \def \binstitute#1{#1}\fi
\ifx \binstitutionaled  \undefined \def \binstitutionaled#1{#1}\fi
\ifx \bctitle  \undefined \def \bctitle#1{#1}\fi
\ifx \beditor  \undefined \def \beditor#1{#1}\fi
\ifx \bpublisher  \undefined \def \bpublisher#1{#1}\fi
\ifx \bbtitle  \undefined \def \bbtitle#1{#1}\fi
\ifx \bedition  \undefined \def \bedition#1{#1}\fi
\ifx \bseriesno  \undefined \def \bseriesno#1{#1}\fi
\ifx \blocation  \undefined \def \blocation#1{#1}\fi
\ifx \bsertitle  \undefined \def \bsertitle#1{#1}\fi
\ifx \bsnm \undefined \def \bsnm#1{#1}\fi
\ifx \bsuffix \undefined \def \bsuffix#1{#1}\fi
\ifx \bparticle \undefined \def \bparticle#1{#1}\fi
\ifx \barticle \undefined \def \barticle#1{#1}\fi
\bibcommenthead
\ifx \bconfdate \undefined \def \bconfdate #1{#1}\fi
\ifx \botherref \undefined \def \botherref #1{#1}\fi
\ifx \url \undefined \def \url#1{\textsf{#1}}\fi
\ifx \bchapter \undefined \def \bchapter#1{#1}\fi
\ifx \bbook \undefined \def \bbook#1{#1}\fi
\ifx \bcomment \undefined \def \bcomment#1{#1}\fi
\ifx \oauthor \undefined \def \oauthor#1{#1}\fi
\ifx \citeauthoryear \undefined \def \citeauthoryear#1{#1}\fi
\ifx \endbibitem  \undefined \def \endbibitem {}\fi
\ifx \bconflocation  \undefined \def \bconflocation#1{#1}\fi
\ifx \arxivurl  \undefined \def \arxivurl#1{\textsf{#1}}\fi
\csname PreBibitemsHook\endcsname

\bibitem[\protect\citeauthoryear{Randall and Sundrum}{1999a}]{Randall1999}
\begin{barticle}
\bauthor{\bsnm{Randall}, \binits{L.}},
\bauthor{\bsnm{Sundrum}, \binits{R.}}:
\batitle{An alternative to compactification}.
\bjtitle{Physical Review Letters}
\bvolume{83},
\bfpage{4690}--\blpage{4693}
(\byear{1999})
\doiurl{10.1103/PhysRevLett.83.4690}
\end{barticle}
\endbibitem

\bibitem[\protect\citeauthoryear{Randall and Sundrum}{1999b}]{Randall1999a}
\begin{barticle}
\bauthor{\bsnm{Randall}, \binits{L.}},
\bauthor{\bsnm{Sundrum}, \binits{R.}}:
\batitle{Large mass hierarchy from a small extra dimension}.
\bjtitle{Physical Review Letters}
\bvolume{83},
\bfpage{3370}--\blpage{3373}
(\byear{1999})
\doiurl{10.1103/PhysRevLett.83.3370}
\end{barticle}
\endbibitem

\bibitem[\protect\citeauthoryear{Bajc and Gabadadze}{2000}]{Bajc2000}
\begin{barticle}
\bauthor{\bsnm{Bajc}, \binits{B.}},
\bauthor{\bsnm{Gabadadze}, \binits{G.}}:
\batitle{Localization of matter and cosmological constant on a brane in anti de
  sitter space}.
\bjtitle{Physics Letters B}
\bvolume{474},
\bfpage{282}--\blpage{291}
(\byear{2000})
\doiurl{10.1016/S0370-2693(00)00055-1}
\end{barticle}
\endbibitem

\bibitem[\protect\citeauthoryear{Jones et~al.}{2013}]{Jones2013}
\begin{barticle}
\bauthor{\bsnm{Jones}, \binits{P.}},
\bauthor{\bsnm{Muñoz}, \binits{G.}},
\bauthor{\bsnm{Singleton}, \binits{D.}},
\bauthor{\bsnm{Triyanta}}:
\batitle{Field localization and the nambu–jona-lasinio mass generation
  mechanism in an alternative five-dimensional brane model}.
\bjtitle{Physical Review D}
\bvolume{88},
\bfpage{025048}
(\byear{2013})
\doiurl{10.1103/PhysRevD.88.025048}
\end{barticle}
\endbibitem

\bibitem[\protect\citeauthoryear{Wulandari et~al.}{2017}]{Wulandari2017}
\begin{botherref}
\oauthor{\bsnm{Wulandari}, \binits{D.}},
\oauthor{\bsnm{Triyanta}},
\oauthor{\bsnm{Kosasih}, \binits{J.S.}},
\oauthor{\bsnm{Singleton}, \binits{D.}},
\oauthor{\bsnm{Jones}, \binits{P.}}:
Localization of interacting fields in five-dimensional braneworld models.
International Journal of Modern Physics A
\textbf{32}
(2017)
\doiurl{10.1142/S0217751X17501913}
\end{botherref}
\endbibitem

\bibitem[\protect\citeauthoryear{Wulandari et~al.}{2019}]{Wulandari2019}
\begin{barticle}
\bauthor{\bsnm{Wulandari}, \binits{D.}},
\bauthor{\bsnm{Triyanta}},
\bauthor{\bsnm{Kosasih}, \binits{J.S.}},
\bauthor{\bsnm{Singleton}, \binits{D.}}:
\batitle{The field localization of yukawa interaction in a modified
  randall-sundrum model}.
\bjtitle{Journal of Physics: Conference Series}
\bvolume{1204},
\bfpage{012044}
(\byear{2019})
\doiurl{10.1088/1742-6596/1204/1/012044}
\end{barticle}
\endbibitem

\bibitem[\protect\citeauthoryear{Liu}{2018}]{Liu2018a}
\begin{botherref}
\oauthor{\bsnm{Liu}, \binits{Y.X.}}:
Introduction to extra dimensions and thick braneworlds.
Memorial Volume for Yi-shi Duan,
211--275
(2018)
\doiurl{10.1142/9789813237278_0008}
\end{botherref}
\endbibitem

\bibitem[\protect\citeauthoryear{Liu et~al.}{2008}]{Liu2008}
\begin{barticle}
\bauthor{\bsnm{Liu}, \binits{Y.-X.}},
\bauthor{\bsnm{Zhang}, \binits{X.-H.}},
\bauthor{\bsnm{Zhang}, \binits{L.-D.}},
\bauthor{\bsnm{Duan}, \binits{Y.-S.}}:
\batitle{Localization of matters on pure geometrical thick branes}.
\bjtitle{Journal of High Energy Physics}
\bvolume{2008},
\bfpage{067}--\blpage{067}
(\byear{2008})
\doiurl{10.1088/1126-6708/2008/02/067}
\end{barticle}
\endbibitem

\bibitem[\protect\citeauthoryear{Guo et~al.}{2013}]{Guo2013}
\begin{barticle}
\bauthor{\bsnm{Guo}, \binits{H.}},
\bauthor{\bsnm{Herrera-Aguilar}, \binits{A.}},
\bauthor{\bsnm{Liu}, \binits{Y.X.}},
\bauthor{\bsnm{Malagón-Morejón}, \binits{D.}},
\bauthor{\bsnm{Mora-Luna}, \binits{R.R.}}:
\batitle{Localization of bulk matter fields, the hierarchy problem and
  corrections to coulomb's law on a pure de sitter thick braneworld}.
\bjtitle{Physical Review D}
\bvolume{87},
\bfpage{1}--\blpage{21}
(\byear{2013})
\doiurl{10.1103/PhysRevD.87.095011}
\end{barticle}
\endbibitem

\bibitem[\protect\citeauthoryear{Herrera-Aguilar
  et~al.}{2010}]{Herrera-Aguilar2010a}
\begin{botherref}
\oauthor{\bsnm{Herrera-Aguilar}, \binits{A.}},
\oauthor{\bsnm{Malagón-Morejón}, \binits{D.}},
\oauthor{\bsnm{Mora-Luna}, \binits{R.R.}}:
Localization of gravity on a de sitter thick braneworld without scalar fields.
Journal of High Energy Physics
\textbf{2010}
(2010)
\doiurl{10.1007/JHEP11(2010)015}
\end{botherref}
\endbibitem

\bibitem[\protect\citeauthoryear{Liang and Duan}{2009}]{Liang2009}
\begin{barticle}
\bauthor{\bsnm{Liang}, \binits{J.}},
\bauthor{\bsnm{Duan}, \binits{Y.-S.}}:
\batitle{Localization of matters on thick branes}.
\bjtitle{Physics Letters B}
\bvolume{678},
\bfpage{491}--\blpage{496}
(\byear{2009})
\doiurl{10.1016/j.physletb.2009.06.067}
\end{barticle}
\endbibitem

\bibitem[\protect\citeauthoryear{Bazeia et~al.}{2009}]{Bazeia2009}
\begin{barticle}
\bauthor{\bsnm{Bazeia}, \binits{D.}},
\bauthor{\bsnm{Gomes}, \binits{A.R.}},
\bauthor{\bsnm{Losano}, \binits{L.}}:
\batitle{Gravity localization on thick branes: A numerical approach}.
\bjtitle{International Journal of Modern Physics A}
\bvolume{24},
\bfpage{1135}--\blpage{1160}
(\byear{2009})
\doiurl{10.1142/S0217751X09043067}
\end{barticle}
\endbibitem

\bibitem[\protect\citeauthoryear{Gremm}{2000}]{Gremm2000a}
\begin{barticle}
\bauthor{\bsnm{Gremm}, \binits{M.}}:
\batitle{Four-dimensional gravity on a thick domain wall}.
\bjtitle{Physics Letters B}
\bvolume{478},
\bfpage{434}--\blpage{438}
(\byear{2000})
\doiurl{10.1016/S0370-2693(00)00303-8}
\end{barticle}
\endbibitem

\bibitem[\protect\citeauthoryear{Geng and Lü}{2016}]{Geng2016}
\begin{botherref}
\oauthor{\bsnm{Geng}, \binits{W.-J.}},
\oauthor{\bsnm{Lü}, \binits{H.}}:
Einstein-vector gravity, emerging gauge symmetry, and de sitter bounce.
Physical Review D
\textbf{93}
(2016)
\doiurl{10.1103/PhysRevD.93.044035}
\end{botherref}
\endbibitem

\bibitem[\protect\citeauthoryear{Dzhunushaliev and
  Folomeev}{2011}]{Dzhunushaliev2011}
\begin{barticle}
\bauthor{\bsnm{Dzhunushaliev}, \binits{V.}},
\bauthor{\bsnm{Folomeev}, \binits{V.}}:
\batitle{Spinor brane}.
\bjtitle{General Relativity and Gravitation}
\bvolume{43},
\bfpage{1253}--\blpage{1261}
(\byear{2011})
\doiurl{10.1007/s10714-010-1105-2}
\end{barticle}
\endbibitem

\bibitem[\protect\citeauthoryear{Cui and Liu}{2023}]{Cui2023}
\begin{botherref}
\oauthor{\bsnm{Cui}, \binits{Z.-Q.}},
\oauthor{\bsnm{Liu}, \binits{Y.-X.}}:
Spinor walls in five-dimensional warped spacetime.
European Physical Journal C
\textbf{83}
(2023)
\doiurl{10.1140/epjc/s10052-023-11422-0}
\end{botherref}
\endbibitem

\bibitem[\protect\citeauthoryear{Bazeia et~al.}{2014}]{Bazeia2014}
\begin{barticle}
\bauthor{\bsnm{Bazeia}, \binits{D.}},
\bauthor{\bsnm{Lobão}, \binits{A.S.}},
\bauthor{\bsnm{Menezes}, \binits{R.}},
\bauthor{\bsnm{Petrov}, \binits{A.}},
\bauthor{\bsnm{Silva}, \binits{A.J.}}:
\batitle{Braneworld solutions for f(r) models with non-constant curvature}.
\bjtitle{Physics Letters, Section B: Nuclear, Elementary Particle and
  High-Energy Physics}
\bvolume{729},
\bfpage{127}--\blpage{135}
(\byear{2014})
\doiurl{10.1016/j.physletb.2014.01.011}
\end{barticle}
\endbibitem

\bibitem[\protect\citeauthoryear{Bazeia et~al.}{2015}]{Bazeia2015}
\begin{barticle}
\bauthor{\bsnm{Bazeia}, \binits{D.}},
\bauthor{\bsnm{Lobão}, \binits{A.S.}},
\bauthor{\bsnm{Menezes}, \binits{R.}}:
\batitle{Thick brane models in generalized theories of gravity}.
\bjtitle{Physics Letters B}
\bvolume{743},
\bfpage{98}--\blpage{103}
(\byear{2015})
\doiurl{10.1016/j.physletb.2015.02.037}
\end{barticle}
\endbibitem

\bibitem[\protect\citeauthoryear{Gu et~al.}{2017}]{Gu2017}
\begin{botherref}
\oauthor{\bsnm{Gu}, \binits{B.M.}},
\oauthor{\bsnm{Zhang}, \binits{Y.P.}},
\oauthor{\bsnm{Yu}, \binits{H.}},
\oauthor{\bsnm{Liu}, \binits{Y.X.}}:
Full linear perturbations and localization of gravity on f(r, t) brane.
European Physical Journal C
\textbf{77}
(2017)
\doiurl{10.1140/epjc/s10052-017-4666-3}
\end{botherref}
\endbibitem

\bibitem[\protect\citeauthoryear{Rohman and Triyanta}{2021}]{Rohman2021b}
\begin{barticle}
\bauthor{\bsnm{Rohman}, \binits{M.T.}},
\bauthor{\bsnm{Triyanta}}:
\batitle{Localization of scalar field on f(r, t) thick robertson-walker brane}.
\bjtitle{Journal of Physics: Conference Series}
\bvolume{1816},
\bfpage{012058}
(\byear{2021})
\doiurl{10.1088/1742-6596/1816/1/012058}
\end{barticle}
\endbibitem

\bibitem[\protect\citeauthoryear{Bezrukov and
  Shaposhnikov}{2008}]{Bezrukov2008}
\begin{barticle}
\bauthor{\bsnm{Bezrukov}, \binits{F.}},
\bauthor{\bsnm{Shaposhnikov}, \binits{M.}}:
\batitle{The standard model higgs boson as the inflaton}.
\bjtitle{Physics Letters B}
\bvolume{659},
\bfpage{703}--\blpage{706}
(\byear{2008})
\doiurl{10.1016/j.physletb.2007.11.072}
\end{barticle}
\endbibitem

\bibitem[\protect\citeauthoryear{Harko and Lobo}{2014}]{Harko2014}
\begin{barticle}
\bauthor{\bsnm{Harko}, \binits{T.}},
\bauthor{\bsnm{Lobo}, \binits{F.S.N.}}:
\batitle{Generalized curvature-matter couplings in modified gravity}.
\bjtitle{Galaxies}
\bvolume{2},
\bfpage{410}--\blpage{465}
(\byear{2014})
\doiurl{10.3390/galaxies2030410}
\end{barticle}
\endbibitem

\bibitem[\protect\citeauthoryear{Farakos and
  Pasipoularides}{2005}]{Farakos2005}
\begin{barticle}
\bauthor{\bsnm{Farakos}, \binits{K.}},
\bauthor{\bsnm{Pasipoularides}, \binits{P.}}:
\batitle{Gravity-induced instability and gauge field localization}.
\bjtitle{Physics Letters, Section B: Nuclear, Elementary Particle and
  High-Energy Physics}
\bvolume{621},
\bfpage{224}--\blpage{232}
(\byear{2005})
\doiurl{10.1016/j.physletb.2005.06.058}
\end{barticle}
\endbibitem

\bibitem[\protect\citeauthoryear{Farakos and
  Pasipoularides}{2006}]{Farakos2006}
\begin{botherref}
\oauthor{\bsnm{Farakos}, \binits{K.}},
\oauthor{\bsnm{Pasipoularides}, \binits{P.}}:
Second randall-sundrum brane world scenario with a nonminimally coupled bulk
  scalar field.
Physical Review D - Particles, Fields, Gravitation and Cosmology
\textbf{73}
(2006)
\doiurl{10.1103/PhysRevD.73.084012}
\end{botherref}
\endbibitem

\bibitem[\protect\citeauthoryear{Guo et~al.}{2012}]{Guo2012}
\begin{botherref}
\oauthor{\bsnm{Guo}, \binits{H.}},
\oauthor{\bsnm{Liu}, \binits{Y.X.}},
\oauthor{\bsnm{Zhao}, \binits{Z.H.}},
\oauthor{\bsnm{Chen}, \binits{F.W.}}:
Thick branes with a nonminimally coupled bulk-scalar field.
Physical Review D
\textbf{85}
(2012)
\doiurl{10.1103/PhysRevD.85.124033}
\end{botherref}
\endbibitem

\bibitem[\protect\citeauthoryear{Moazzen and Ghalenovi}{2017}]{Moazzen2017}
\begin{barticle}
\bauthor{\bsnm{Moazzen}, \binits{M.}},
\bauthor{\bsnm{Ghalenovi}, \binits{Z.}}:
\batitle{Non-minimally coupled bulk scalar fields in sine–gordon braneworld
  models}.
\bjtitle{Annals of Physics}
\bvolume{385},
\bfpage{70}--\blpage{85}
(\byear{2017})
\doiurl{10.1016/j.aop.2017.08.006}
\end{barticle}
\endbibitem

\bibitem[\protect\citeauthoryear{Zhao et~al.}{2023}]{Zhao2023}
\begin{botherref}
\oauthor{\bsnm{Zhao}, \binits{Z.-H.}},
\oauthor{\bsnm{Xie}, \binits{Q.-Y.}},
\oauthor{\bsnm{Fu}, \binits{C.-E.}},
\oauthor{\bsnm{Zhou}, \binits{X.-N.}}:
Localization of u(1) gauge field by non-minimal coupling with gravity in
  braneworlds.
Journal of Cosmology and Astroparticle Physics
\textbf{2023}
(2023)
\doiurl{10.1088/1475-7516/2023/07/010}
\end{botherref}
\endbibitem

\bibitem[\protect\citeauthoryear{Moreira et~al.}{2023}]{Moreira2023}
\begin{botherref}
\oauthor{\bsnm{Moreira}, \binits{A.R.P.}},
\oauthor{\bsnm{Belchior}, \binits{F.M.}},
\oauthor{\bsnm{Maluf}, \binits{R.V.}},
\oauthor{\bsnm{Almeida}, \binits{C.A.S.}}:
Gauge field localization in branes: coupling to a scalar function and coupling
  to torsion in teleparallel gravity scenario.
European Physical Journal C
\textbf{83}
(2023)
\doiurl{10.1140/epjc/s10052-023-11213-7}
\end{botherref}
\endbibitem

\bibitem[\protect\citeauthoryear{Zhong and Sui}{2024}]{Zhong2024}
\begin{botherref}
\oauthor{\bsnm{Zhong}, \binits{Y.}},
\oauthor{\bsnm{Sui}, \binits{T.-T.}}:
Localization of q-form fields on a de sitter brane in chameleon gravity.
Frontiers of Physics
\textbf{19}
(2024)
\doiurl{10.1007/s11467-023-1361-8}
\end{botherref}
\endbibitem

\bibitem[\protect\citeauthoryear{Zhou and Guo}{2023}]{Zhou2023}
\begin{botherref}
\oauthor{\bsnm{Zhou}, \binits{X.-N.}},
\oauthor{\bsnm{Guo}, \binits{W.-Q.}}:
Localization of gravitino field with non-minimal coupling on a thick brane.
Physica Scripta
\textbf{98}
(2023)
\doiurl{10.1088/1402-4896/ad0527}
\end{botherref}
\endbibitem

\bibitem[\protect\citeauthoryear{DeWolfe et~al.}{2000}]{DeWolfe2000}
\begin{barticle}
\bauthor{\bsnm{DeWolfe}, \binits{O.}},
\bauthor{\bsnm{Freedman}, \binits{D.Z.}},
\bauthor{\bsnm{Gubser}, \binits{S.S.}},
\bauthor{\bsnm{Karch}, \binits{A.}}:
\batitle{Modeling the fifth dimension with scalars and gravity}.
\bjtitle{Physical Review D - Particles, Fields, Gravitation and Cosmology}
(\byear{2000})
\doiurl{10.1103/PhysRevD.62.046008}
\end{barticle}
\endbibitem

\bibitem[\protect\citeauthoryear{Nambu and Jona-Lasinio}{1961}]{Nambu1961}
\begin{barticle}
\bauthor{\bsnm{Nambu}, \binits{Y.}},
\bauthor{\bsnm{Jona-Lasinio}, \binits{G.}}:
\batitle{Dynamical model of elementary particles based on an analogy with
  superconductivity. i}.
\bjtitle{Physical Review}
\bvolume{122},
\bfpage{345}--\blpage{358}
(\byear{1961})
\doiurl{10.1103/PhysRev.122.345}
\end{barticle}
\endbibitem

\bibitem[\protect\citeauthoryear{Cruz et~al.}{2016}]{Cruz2016}
\begin{barticle}
\bauthor{\bsnm{Cruz}, \binits{W.T.}},
\bauthor{\bsnm{Maluf}, \binits{R.V.}},
\bauthor{\bsnm{Sousa}, \binits{L.J.S.}},
\bauthor{\bsnm{Almeida}, \binits{C.A.S.}}:
\batitle{Gravity localization in sine-gordon braneworlds}.
\bjtitle{Annals of Physics}
\bvolume{364},
\bfpage{25}--\blpage{34}
(\byear{2016})
\doiurl{10.1016/j.aop.2015.10.016}
\end{barticle}
\endbibitem

\bibitem[\protect\citeauthoryear{Davoudiasl et~al.}{2000}]{Davoudiasl2000}
\begin{barticle}
\bauthor{\bsnm{Davoudiasl}, \binits{H.}},
\bauthor{\bsnm{Hewett}, \binits{J.L.}},
\bauthor{\bsnm{Rizzo}, \binits{T.G.}}:
\batitle{Bulk gauge fields in the randall-sundrum model}.
\bjtitle{Physics Letters, Section B: Nuclear, Elementary Particle and
  High-Energy Physics}
\bvolume{473},
\bfpage{43}--\blpage{49}
(\byear{2000})
\doiurl{10.1016/S0370-2693(99)01430-6}
\end{barticle}
\endbibitem

\bibitem[\protect\citeauthoryear{Guerrero and Rodriguez}{2019}]{Guerrero2019}
\begin{barticle}
\bauthor{\bsnm{Guerrero}, \binits{R.}},
\bauthor{\bsnm{Rodriguez}, \binits{R.O.}}:
\batitle{Fermions localization on de sitter branes}.
\bjtitle{Physical Review D}
\bvolume{100},
\bfpage{104038}
(\byear{2019})
\doiurl{10.1103/PhysRevD.100.104038}
\end{barticle}
\endbibitem

\bibitem[\protect\citeauthoryear{Oda}{2000}]{Oda2000}
\begin{barticle}
\bauthor{\bsnm{Oda}, \binits{I.}}:
\batitle{Localization of matters on a string-like defect}.
\bjtitle{Physics Letters, Section B: Nuclear, Elementary Particle and
  High-Energy Physics}
\bvolume{496},
\bfpage{113}--\blpage{121}
(\byear{2000})
\doiurl{10.1016/S0370-2693(00)01284-3}
\end{barticle}
\endbibitem

\end{thebibliography}

\end{document}